\documentclass[12pt, letter]{article}
\usepackage{amssymb}
\usepackage{amsmath}
\usepackage{amssymb}
\usepackage{subcaption}
\usepackage{amsmath, setspace}
\usepackage{amsmath, amssymb, pgfplots, tikz, tabularx}
\usepackage{pgfplots,tikz, verbatim}
\usepackage{fixltx2e}
\usepackage{siunitx}
\usepackage{xcolor}
\usepackage{xr}
\usepackage[version=4]{mhchem}
\usepackage{pgfplots}
\usepackage{a4}
\usepackage[a4paper,left=3cm,right=3cm,top=3.1cm,bottom=3.3cm]{geometry}
\usepackage{todonotes}

\usetikzlibrary{patterns}
\usepgfplotslibrary{external}
\pgfplotsset{compat=newest} 
\usepgfplotslibrary{groupplots}
\usetikzlibrary{matrix,calc}
\DeclareSIUnit{\molar}{M}
\pgfplotsset{compat=newest}
\usepgfplotslibrary{fillbetween}
\newenvironment{proof}[1][Proof]{\textbf{#1.} }{\ \rule{0.5em}{0.5em}}
\usepgfplotslibrary{groupplots}
\usetikzlibrary{pgfplots.groupplots}
\usetikzlibrary{plotmarks}
\usetikzlibrary{patterns}
\usepgfplotslibrary{external}
\pgfplotsset{compat=newest} 
\newcommand{\ci}{\perp\!\!\!\perp}

\input{texdraw}
\usepgfplotslibrary{external}
\usepgfplotslibrary{groupplots}
\newcounter{corollary_counter}
\newcounter{definition_counter}
\newcounter{condition_counter}
\newcounter{lemma_counter}
\newcounter{remark_counter}
\newtheorem{theorem}{Theorem}
\newtheorem{algorithm}[theorem]{Algorithm}
\newtheorem{assumption}{Assumption}

\newtheorem{condition}[condition_counter]{Condition}

\newtheorem{corollary}[corollary_counter]{Corollary}

\newtheorem{definition}[definition_counter]{Definition}

\newtheorem{lemma}[lemma_counter]{Lemma}

\newtheorem{remark}[remark_counter]{Remark}

\renewcommand{\Pr}{\mathbb{P}}

\providecommand{\keywords}[1]{\noindent \textbf{Keywords:} #1 \\}
\providecommand{\jelcodes}[1]{\noindent \textbf{JEL-codes:} #1 \\}

\renewcommand{\Pr}{\mathbb{P}}

\providecommand{\keywords}[1]{\noindent \textbf{Keywords:} #1 \\}
\providecommand{\jelcodes}[1]{\noindent \textbf{JEL-codes:} #1 \\}
\numberwithin{equation}{section}
\begin{document}

\title{\textbf{Nonseparable Sample Selection Models with Censored Selection
Rules}\thanks{
We thank Costas Meghir for providing the data, and the editor Xiaohong Chen,
an associate editor, two anonymous referees, St\'{e}phane Bonhomme, Ivana
Komunjer and Sami Stouli for useful comments.}}
\author{Ivan Fern\'{a}ndez-Val\thanks{%
Boston University.} \quad Aico van Vuuren\thanks{%
University of Gothenburg.} \quad Francis Vella\thanks{%
Georgetown University. }}
\date{\today }
\maketitle

\begin{abstract}
We consider identification and estimation of nonseparable sample selection
models with censored selection rules. We employ a control function approach
and discuss different objects of interest based on (1) local effects
conditional on the control function, and (2) global effects obtained from
integration over ranges of values of the control function. We derive the
conditions for the identification of these different objects and suggest
strategies for estimation. Moreover, we provide the associated asymptotic
theory. These strategies are illustrated in an empirical investigation of
the determinants of female wages in the United Kingdom. \thispagestyle{empty}
\end{abstract}


\jelcodes{C14, C21, C24}

\newpage \setcounter{page}{1}

\section{Introduction}

This paper considers a nonseparable sample selection model with a censored
selection rule. The most common example is a selection rule with censoring
at zero, also referred to in the parametric setting as tobit type 3,
although other forms of censored selection rules are permissible. A leading
empirical example is estimating the determinants of wages when workers
report working hours rather than the binary work/not work decision. We
account for selection via an appropriately constructed control function and
propose a three-step estimation procedure which first employs distribution
regression to compute the appropriate control function. The second step is
estimated by least squares, distribution or quantile regression employing
the estimated control function as an additional conditioning variable. The
estimands of interest are obtained in the third step as functionals of the
second step and control function estimates.

Our paper contributes to the growing literatures on nonseparable models with
endogeneity (see, for example, Chesher 2003, Ma and Koenker 2006, Florens et
al. 2008, Imbens and Newey 2009, Jun 2009 and Masten and Torgovitsky, 2016)
and nonseparable sample selection models (for example, Newey 2007). An
important focus of our paper is on the identification and estimation of
local effects. 
This local approach to identification is popular in many contexts (see, for
example, Chesher 2003, and Heckman and Vytlacil 2005). We show that for any
population observation with positive probability of being selected,
selection is irrelevant for the distribution of the outcome variable
conditional on the control function. Hence, our local objects of interest
are identified for the whole population and not just the selected one. We
can also estimate global objects by integrating over the distribution of the
control function in the selected population. However, global objects are
only nonparametrically identified under support assumptions for the
explanatory variables which may be difficult to satisfy in empirical
applications. Accordingly, we also consider global effects \textquotedblleft
for the treated\textquotedblright\ that are identified under weaker
assumptions.

We provide estimators of the global and local effects and their associated
asymptotic theory. These are based on semiparametric specifications of the
first and second steps of the estimation procedure. The semiparametric
structure overcomes the curse of dimensionality and the reliance on support
assumptions required to nonparametrically identify the global effects (Newey
and Stouli, 2018, Chernozhukov et al., 2020). It can also be used as the
basis for sieve methods by increasing the dimension of the specification
with the sample size (Chen, 2007). 
We recommend that the estimates of the global effects are supplemented with 
bounds that only require the semi-parametric structure be correctly
specified within the support, whereas the point estimators rely on
extrapolation outside the support. We provide estimates for these bounds.

Our paper is closest to Newey (2007) who uses a control function approach to
correct for sample selection in a nonseparable model.\label{page:newey}
Nevertheless, there are several differences. First, Newey (2007) examines a
binary selection rule whereas we consider a censored selection rule. Second,
in addition to considering global parameters for the selected population, as
is done in Newey (2007), we also consider local parameters for the entire
population and identification of the effects \textquotedblleft for the
treated\textquotedblright . Third, we follow the suggestion by Newey (2007)
to construct bounds for the global parameters for the entire population.
Finally, we provide estimation and inference results in addition to
addressing identification issues.

Our paper is also related to Caetano et al. (2016), which studied a
nonseparable model with an endogenous regressor featuring a mass point at
some known value. This includes censoring as a special case. They build on
Caetano (2015) by developing a test for the validity of the control function
approach. These papers, and ours, are motivated by the Imbens and Newey
(2009) approach although each paper provides a treatment of important but
different issues. They focus on specification testing, whereas we analyze
the estimation and identification of local effects in the presence of sample
selection. We also consider the use of bounds for the global effects. Thus,
we view our paper as complementary to Caetano et al. (2016).

This paper also contributes to the literature on quantile selection models.
Arellano and Bonhomme (2017) addressed selection by modeling the copula of
the error terms in the outcome and selection equations. The most important
distinction to this paper is that they consider the conventional binary,
rather than a censored, selection equation. Thus, we require more
information about the selection process. However, this has the advantage
that one can consider local effects conditional on the control function
which are identified under weaker conditions.

We acknowledge that the binary selection model is more frequently employed
in empirical work than the censored selection model. However, this partially
reflects the remarkable popularity of the Heckman (1979) two-step procedure.
In fact, in many empirical investigations, researchers dichotomize censored
selection variables in order to employ the Heckman procedure. Examples of
such commonly encountered censoring selection variables include unemployment
duration, training program length and the magnitude or the length of the
welfare benefit receipt. In the parametric setting, there are no substantial
benefits in using the censored, rather than the dichotomized form of the
selection variable other than that the selection variable can appear in the
outcome equation as a regressor and the variation in this variable provides
an additional form of identification. However, in the nonseparable setting,
a censored selection rule enables inference for both the selected and
non-selected populations and the investigation of local effects. Finally, we
illustrate that the additional information implicit in the censored
selection setting can be exploited to reduce the bounds for the global
effects. Thus, it could be argued that the censored selection approach
should be employed when possible.\label{page::censored_better}

The following section outlines the model and some related literature.
Section \ref{sec:identification} defines the control function and provides
identification results regarding the objects of interest in the model.
Section \ref{sec:estimation} provides estimators of these objects and
discusses inference. Section \ref{sec:empirical_example} illustrates some of
our estimands focusing on the determinants of female wages in the United
Kingdom.

\section{Model}

\label{sec:model}

The model has the following structure: 
\begin{eqnarray}
Y &=&g(X,Z_{1},\varepsilon )\text{ if }C>0,  \label{1} \\
C &=&\max \left( h(Z,\eta \right) ,0),  \label{2}
\end{eqnarray}%
where $Y$ and $C$ are observable random variables, and $X$, $Z_{1}$ and $%
Z:=(Z_{1},Z_{2})$ are vectors of observable explanatory variables. The
variables included in $X$ are a subset of those included in $Z_{2}$, i.e. $X
\subseteq Z_2$. We do not need to impose an exclusion restriction on $Z_{2}$
with respect to the elements of $X$, although our nonparametric
identification assumptions will be more plausible with such a restriction.
We separate $X$ from $Z_{1}$ in \eqref{1} to distinguish between the
variables of interest or treatments, $X$, from the rest of the explanatory
variables, $Z_{1}$, which play the role of controls. The functions $g$ and $h
$ are unknown and $\varepsilon $ and $\eta $ are a vector and a scalar of
potentially mutually dependent unobservables, respectively. We shall impose
restrictions on the stochastic properties of these unobservables. The
primary objective is to estimate functionals related to $g$ noting that $Y$
is only observed when $C$ is above some known threshold normalized to be
zero. The non observability of $Y $ for specific values of $C$ induces the
possibility of selection bias. We refer to \eqref{1} as the outcome equation
and \eqref{2} as the selection equation.

The model is a nonparametric and nonseparable representation of the tobit
type-3 model and is a variant of the Heckman (1979) selection model. It was
initially examined in a fully parametric setting, imposing additivity and
normality, and estimated by maximum likelihood (see Amemiya, 1978, 1979).
Vella (1993) provided a two-step estimator based on estimating the
generalized residual from the selection equation and including it as a
control function in the outcome equation. Honor\'{e} et al. (1997), Chen
(1997) and Lee and Vella (2006) provide semi-parametric estimators for this
model.

The model can be extended in several directions. For example, the selection
variable $C$ could be censored in a number of ways provided that there are
some region(s) for which it is continuously observed. This allows for top,
middle and/or bottom censoring. In addition, although we do not explicitly
consider it here, our approach is applicable when the outcome variable $Y$
is also censored. For example: 
\begin{equation*}
Y=\max (g(X,Z_{1},\varepsilon ),0)\text{ if }C>0.
\end{equation*}%
It is also applicable with random censoring in the selection equation. That
is, where \eqref{2} is replaced with:\label{insert::censored_model} 
\begin{equation*}
C=\max \left( h(Z,\eta \right) ,\varsigma )
\end{equation*}%
where $\varsigma $ is a random variable which is independent of ($%
\varepsilon ,\eta $) conditional on $Z.\footnote{%
The appropriate control function can be obtained via the Kaplan Meier
approach conditional on $Z$. The proof is available from the authors upon
request. We are grateful to the Associate Editor for this suggestion.}$ The
model can also be extended to include $C$ in the outcome equation as an
explanatory variable provided that there is an exclusion restriction in $%
Z_{2}$ with respect to $X$. This extension, which corresponds to the
triangular system of Imbens and Newey (2009) with censoring in the
first-stage equation and selection in the second, is not considered here as
it is not relevant for our empirical application. However, note that the
Imbens and Newey (2009) estimator does not apply to this case as they do not
consider the selection model.

We highlighted above that we follow a local approach to identification such
as proposed by Heckman and Vytlacil (2005) who consider a binary
treatment/selection rule and a separable selection equation. While our focus
is also, in part, on local effects, our model differs with respect to the
selection rule and the possible presence of nonseparability.

\section{Identification of objects of interest}

\label{sec:identification}

We account for selection bias via the use of an appropriately constructed
control function. We establish the existence of such a function for this
model and then define some objects of interest incorporated in (\ref{1})-(%
\ref{2}).

Let $\perp\!\!\!\perp$ denote stochastic independence. We begin with the
following assumption:

\begin{assumption}[Control Function]
\label{assumption:cv} $\left( \varepsilon ,\eta \right) \perp\!\!\!\perp Z $%
, $\eta$ is a continuously distributed random variable with strictly
increasing CDF on the support of $\eta$, and $t \mapsto h(Z,t)$ is strictly
increasing a.s.
\end{assumption}
This assumption allows for endogeneity between $(X,Z_1)$ and $\varepsilon $
in the selected population with $C>0$, since, in general, $\varepsilon $ and 
$\eta $ are dependent, \emph{i.e.} $\varepsilon \not \ci(X,Z_1)\mid C>0$. It
allows for a non-monotonic relationship between $\varepsilon $ and $C$
because $\varepsilon $ and $\eta $ are allowed to be non-monotonically
dependent. Under Assumption \ref{assumption:cv}, we can normalize the
distribution of $\eta $ to be uniform on $[0,1]$ without loss of generality
(Matzkin, 2003).\footnote{%
Indeed if $t\mapsto h(z,t)$ is strictly increasing, and $\eta $ is
continuously distributed with $\eta \sim F_{\eta }$, then $\tilde{h}(z,%
\tilde{\eta})=h(z,F_{\eta }(\tilde{\eta}))$ is such that $t\mapsto \tilde{h}%
(z,t)$ is strictly increasing and $\tilde{\eta}\sim U(0,1)$.}

The following lemma shows the existence of a control function for the
selected population in this setting. That is, there is a function of the
observable data such that the unobservable component is independent of the
explanatory variables in the outcome equation for the selected population
conditional on this function. Let $V:=F_{C|Z}(C\mid Z)$ where $F_{C|Z}(\cdot
\mid z)$ denotes the CDF of $C$ conditional on $Z=z$.

\begin{lemma}[Existence of Control Function]
\label{lem:cv} Under the model in (\ref{1})-(\ref{2}) and Assumption \ref%
{assumption:cv}: 
\begin{equation*}
\varepsilon \perp \!\!\!\perp Z\mid V,C>0.
\end{equation*}
\end{lemma}
All proofs are provided in the Appendix. The intuition behind Lemma \ref%
{lem:cv} is based on three observations. First, $V=\eta$ when $C>0,$ so that
conditioning on $V$ is identical to conditioning on $\eta$ in the selected
population. 
Second, conditioning on $Z$ and $\eta $ makes the selection, \emph{i.e.} $%
C>0 $, deterministic. Therefore, the distribution of $\varepsilon $,
conditional on $Z$ and $\eta $, does not depend on the condition that $C>0$.
The final observation, namely our assumption that $(\varepsilon,\eta) \perp
\!\!\!\perp Z,$ is sufficient to prove the Lemma.

We consider two classes of objects of interest. These are: (1) local effects
conditional on the value of the control function, and (2) global effects
based on integration over the control function.

\subsection{Local effects}

\label{ss:local}

We consider local effects on $Y$ for given values of $X$ conditional on the
control function $V$. Let $\mathcal{Z}$, $\mathcal{X}$, $\mathcal{Z}_1$, and 
$\mathcal{V}$ denote the marginal supports of $Z$, $X$, $Z_1$ and $V$ in the
selected population, respectively. We define $\mathcal{XZ}_1\mathcal{V}$ as
the joint support of $X$, $Z_1$ and $V$ in the selected population:


\begin{definition}[Identification set]
\label{assumption:relevance} Define 
\begin{equation*}
\mathcal{XZ}_1\mathcal{V}:=\left\{ (x,z_1,v)\in \mathcal{X}\times \mathcal{Z}%
_1 \times \mathcal{V}:h(z,v)>0,z\in \mathcal{Z}(x,z_1)\right\},
\end{equation*}
where $\mathcal{Z}(x,z_1)=\{z\in \mathcal{Z}:(x,z_1)\subseteq z\},$ \emph{%
i.e.} the set of values of $Z$ with the components $X=x$ and $Z_1=z_1$. %
%
\end{definition}

Depending on the values of $(X,Z_1,\eta)$, we can classify the units of
observation into 3 groups: (1) always selected units when $h(z,t)>0$ for all 
$z\in \mathcal{Z}(x,z_1)$, (2) switchers when $h(z,t)>0$ for some $z\in 
\mathcal{Z}(x,z_1)$ and $h(z,t)\leq 0$ for some $z\in \mathcal{Z}(x,z_1)$,
and (3) never selected units when $h(z,t)\leq 0$ for all $z\in \mathcal{Z}%
(x,z_1)$. The set $\mathcal{XZ}_1\mathcal{V}$ only includes always selected
units and switchers, \emph{i.e.} units with $(X,Z_1,V)$ such that they are
observed for some values of $Z$. When $X=Z_2$, there are no switchers
because the set $\mathcal{Z}(x)$ is a singleton. Otherwise the size of the
set $\mathcal{XZ}_1\mathcal{V}$ increases with the support of the excluded
variables and their strength in the selection equation.


We now define the local average structural function.

\begin{definition}[LASF]
\label{def:condition_expectation_pop} The local average structural function
(LASF) at $(x,z_1,v)$ is:%
\begin{equation*}
\mu (x,z_1,v)=\mathbb{E}(g(x,z_1,\varepsilon ) \mid V=v).
\end{equation*}
\end{definition}
The LASF gives the expected value of the potential outcome $%
g(x,z_1,\varepsilon ) $ obtained by fixing $(X,Z_1)$ at $(x,z_1)$
conditional on $V=v$ for the entire population. It is useful for measuring
the effect of $X$ on the mean of $Y$. For example, the average treatment
effect of changing $X$ from $x_{0}$ to $x_{1}$ at $Z_1 = z_1$ conditional on 
$V=v$ is: 
\begin{equation*}
\mu (x_{1},z_1,v)-\mu (x_{0},z_1,v).
\end{equation*}
%
%
%
%
%
%
%
%
%
%
%
%
%
%
%
%
The following result shows that $\mu (x,z_1,v)$ is identified for all $%
(x,z_1,v)\in \mathcal{XZ}_1\mathcal{V}$. 

\begin{theorem}[Identification of LASF]
\label{lem:ident_d_star} Under the model (\ref{1})-(\ref{2}), Assumption \ref%
{assumption:cv} and $\mathbb{E} |Y| < \infty$, for $(x,z_1,v)\in \mathcal{XZ}%
_1\mathcal{V}$, 
\begin{equation}
\mu (x,z_1,v)= \mathbb{E}(Y \mid X=x,Z_1 = z_1,V=v,C>0).
\label{eq:asf_identified}
\end{equation}
\end{theorem}
According to Theorem \ref{lem:ident_d_star}, the LASF is identical to the
expected value of $Y$ conditional on $(X,Z_1,V)=(x,z_1,v)$ in the selected
population. The proof is based on Assumption \ref{assumption:cv} that allows
for the LASF to be conditional on $(Z,V)=(z,v)$. Since $(x,z_1,v)\in 
\mathcal{XZ}_1\mathcal{V}$, there is a $z\in \mathcal{Z}(x,z_1)$ such that $%
h(z,v)>0$, the expected value of $g(X,Z_1,\varepsilon )$ conditional on $V=v$
for the entire population, \emph{i.e.} the LASF, is the same as that
expected value of $Y$ for the selected population conditional on $(X,Z_1,V)
= (x,z_1,v)$.


We can consider 
the average derivative of $g(x,z_1,\varepsilon )$ with respect to $x$
conditional on the control function when $X$ is continuous and $x\mapsto
g(x,z_1,\varepsilon)$ is differentiable almost surely.

\begin{definition}[LADF]
\label{def:local_average_derivative_pop} The local average derivative
function (LADF) at $(x,z_1,v)$ is: 
\begin{equation}
\delta (x,z_1,v)=\mathbb{E}[\partial _{x}g(x,z_1,\varepsilon )\mid V=v],\ \
\ \partial _{x}:=\partial /\partial x.  \label{eq:average_derivative}
\end{equation}
\end{definition}
The LADF is the first-order derivative of the LASF with respect to $x$,
provided that we can interchange differentiation and integration in (\ref%
{eq:average_derivative}). This is made formal in the next corollary which
shows that the LADF is identified for all $(x,z_1,v)\in \mathcal{XZ}_1%
\mathcal{V}$.

\begin{corollary}[Identification of LADF]
\label{theorem:local_average_derivative} Assume that for all $(x,z_1) \in 
\mathcal{X}\times\mathcal{Z}_1$, $x \mapsto g(x,z_1,\varepsilon )$ is
continuously differentiable a.s., $\mathbb{E}[|g(x,z_1,\varepsilon
)|]<\infty $, and $\mathbb{E}[|\partial _{x}g(x,$ $z_1,\varepsilon
)|]<\infty $. Under the conditions of Theorem \ref{lem:ident_d_star}, for $%
(x,z_1,v)\in \mathcal{XZ}_1\mathcal{V}$, 
\begin{equation*}
\delta (x,z_1,v)= \partial_x \mu (x,z_1,v)=\partial _{x}\mathbb{E}(Y \mid
X=x,Z_1=z_1,V=v,C>0).
\end{equation*}
\end{corollary}
%
%
%
%
The local effects extend in a straightforward manner to distributions and
quantiles.

\begin{definition}[LDSF and LQSF]
The local distribution structural function (LDSF) at $(y,x,z_1,v)$ is: 
\begin{equation*}
G(y,x,z_1,v)=\mathbb{E}[1\left\{ g(x,z_1,\varepsilon )\leq y\right\} \mid
V=v].
\end{equation*}
The local quantile structural function (LQSF) at $(\tau,x,z_1,v)$ is: 
\begin{equation*}
q(\tau ,x,z_1,v):=\inf \{y\in \mathbb{R}:G(y,x,z_1,v)\geq \tau \}.
\end{equation*}
\end{definition}
The LDSF is the distribution function of the potential outcome $%
g(x,z_1,\varepsilon )$ conditional on the value of the control function for
the entire population. The LQSF is the left-inverse function of $y\mapsto
G(y,x,z_1,v)$ and corresponds to the quantiles of $g(x,z_1,\varepsilon )$.
The differences of the LQSF across levels of $x$ correspond to quantile
treatment effects conditional on $V$ for the entire population. For example,
the $\tau $-quantile treatment effect of changing $X$ from $x_{0}$ to $x_{1}$
is: 
\begin{equation*}
q(\tau ,x_{1},z_1,v)-q(\tau ,x_{0},z_1,v).
\end{equation*}%
%
%
%
%
%
%
%
%
%
%
%
%
%
%
%
%
%
%
%
The identification of the LDSF follows by the same argument as the
identification of the LASF, replacing $g(x,z_1,\varepsilon )$ (as in
Definition \ref{def:condition_expectation_pop}) by $1\left\{
g(x,z_1,\varepsilon )\leq y\right\} $ and $Y$ (as in equation (\ref%
{eq:asf_identified})) by $1\{Y\leq y\}$. Thus, under Assumption \ref%
{assumption:cv}, for $(x,z_1,v)\in \mathcal{XZ}_1\mathcal{V}$, 
\begin{equation*}
\mathbb{E}[1\left\{ g(x,z_1,\varepsilon )\leq y\right\} \mid
V=v]=F_{Y|X,Z_1,V,C>0}(y\mid x,z_1,v).
\end{equation*}%
The LQSF is then identified by the left-inverse function of $y\mapsto
F_{Y|X,Z_1,V,C>0}(y\mid x,z_1,v)$, the conditional quantile function $\tau
\mapsto \mathbb{Q}_{Y}[\tau \mid X=x,Z_1=z_1,V=v,C>0]$, \emph{i.e.}, for $%
(x,z_1,v)\in \mathcal{XZ}_1\mathcal{V}$, 
\begin{equation*}
q(\tau ,x,z_1,v)=\mathbb{Q}_{Y}[\tau \mid X=x,Z_1=z_1,V=v,C>0].
\end{equation*}%
We also consider the derivative of $q(\tau ,x,z_1,v)$ with respect to $x$
and call it the local quantile derivative function (LQDF). This object
corresponds to the average derivative of $g(x,z_1,\varepsilon )$ with
respect to $x$ at the quantile $q(\tau ,x,z_1,v)$ conditional on $V=v$ under
suitable regularity conditions; see Hoderlein and Mammen (2011). Thus, for $%
(\tau ,x,z_1,v)\in \lbrack 0,1]\times \mathcal{XZ}_1\mathcal{V},$ 
\begin{equation*}
\delta _{\tau }(x,z_1,v):=\partial _{x}q(\tau ,x,z_1,v)=\mathbb{E}[\partial
_{x}g(x,z_1,\varepsilon )\mid V=v,g(x,z_1,\varepsilon )=q(\tau ,x,z_1,v)].
\end{equation*}%
By an analogous argument to Corollary \ref{theorem:local_average_derivative}%
, the LQDF is identified at $(\tau ,x,z_1,v)\in \lbrack 0,1]\times \mathcal{%
XZ}_1\mathcal{V}$ by: 
\begin{equation*}
\delta _{\tau }(x,z_1,v)=\partial _{x}\mathbb{Q}_{Y}[\tau \mid
X=x,Z_1=z_1,V=v,C>0],
\end{equation*}%
provided that $x\mapsto \mathbb{Q}_{Y}[\tau \mid X=x,Z_1=z_1,V=v,C>0]$ is
differentiable and other regularity conditions hold. 

%
%

\begin{remark}[Exclusion restrictions]
\label{er} The identification of local effects does not explicitly require
exclusion restrictions in $Z_2$ with respect to $X$ although the size of the
identification set $\mathcal{XZ}_1\mathcal{V}$ depends on such restrictions.
For example, dropping $Z_1$, if $h(z,\eta )=z+\Phi ^{-1}(\eta )$ where $\Phi 
$ is the standard normal distribution and $X=Z$, then $\mathcal{XV}%
=\{(x,v)\in \mathcal{X}\times \mathcal{V}]:x> -\Phi ^{-1}(v)\}\subset 
\mathcal{X}\times \mathcal{V}$; whereas if $h(z,\eta )=x+\tilde z+\Phi
^{-1}(\eta )$ for $Z=(X,\tilde Z)$, then $\mathcal{XV}=\{(x,v)\in \mathcal{X}%
\times \mathcal{V}:x> -\Phi ^{-1}(v)-\tilde z,\tilde z \in \mathcal{Z}(x)\}$%
, such that $\mathcal{XV}=\mathcal{X}\times \mathcal{V}$ if $\tilde Z$ is
independent of $X$ and supported in $\mathbb{R}$.
\end{remark}

%

\subsection{Global effects}

\label{ss:integration}

We expand our set of estimands by examining the global counterparts of the
local effects obtained by integration over the control function and the
observed controls in the selected population. A typical global effect at $%
x\in \mathcal{X}$ is: 
\begin{equation}
\theta_S (x)=\int \theta (x,z_1,v)dF_{Z_1,V \mid C>0}(z_1,v),  \label{ge}
\end{equation}%
where $\theta (x,z_1,v)$ can be any of the local objects defined above and $%
F_{Z_1,V \mid C>0} $ is the joint distribution of $(Z_1,V)$ in the selected
population. The identification of $\theta_S (x)$ requires identification of $%
\theta (x,z_1,v)$ over $\mathcal{Z}_1\mathcal{V}$, the support of $(Z_1,V)$
in the selected population.

For example, the average structural function (ASF): 
\begin{equation*}
\mu _{S}(x):=\mathbb{E}[g(x,Z_1,\varepsilon )\mid C>0],
\end{equation*}%
gives the average of the potential outcome $g(x,Z_1,\varepsilon )$ in the
selected population. By the law of iterated expectations, this is a special
case of the global effect \eqref{ge} with $\theta (x,z_1,v)=\mu (x,z_1,v)$,
the LASF. The average treatment effect of changing $X$ from $x_{0}$ to $%
x_{1} $ in the selected population is: 
\begin{equation}
\mu _{S}(x_{1})-\mu _{S}(x_{0}).  \label{eq:estimated}
\end{equation}%
Similarly, one can consider the distribution structural function (DSF) in
the selected population as in Newey (2007), \emph{i.e:} 
\begin{equation*}
G_{S}(y,x):=\mathbb{E}[1\{g(x,Z_1,\varepsilon )\leq y\}\mid C>0],
\end{equation*}%
which gives the distribution of the potential outcome $g(x,Z_1,\varepsilon )$
at $y$ in the selected population. This is also a special case of the global
effect \eqref{ge} with $\theta (x,z_1,v)=G(y,x,z_1,v)$. We can then
construct the quantile structural function (QSF) in the selected population
as the left-inverse of $y\mapsto G_{S}(y,x)$, that is: 
\begin{equation*}
q_{S}(\tau ,x):=\inf \{y\in \mathbb{R}:G_{S}(y,x)\geq \tau \}.
\end{equation*}%
The QSF gives the quantiles of $g(x,Z_1,\varepsilon )$. Unlike $G_{S}(y,x)$, 
$q_{S}(\tau ,x)$ cannot be obtained by integration of the corresponding
local effect, $q(\tau ,x,z_1,v)$, because we cannot interchange quantiles
and expectations. The $\tau $-quantile treatment effect of changing $X$ from 
$x_{0}$ to $x_{1}$ in the selected population is: 
\begin{equation*}
q_{S}(\tau ,x_{1})-q_{S}(\tau ,x_{0}).
\end{equation*}%
The global counterparts of the LADF and LQSF are obtained by taking the
derivatives of $\mu _{S}(x)$ and $q_{S}(\tau ,x)$ with respect to $x$.


As in Newey (2007), the identification of the global effects in the selected
population requires a condition on the support of the control function. Let $%
\mathcal{Z}_1\mathcal{V}(x)$ denote the support of $(Z_1,V)$ conditional on $%
X=x$, \emph{i.e.} $\mathcal{Z}_1\mathcal{V}(x) := \{(z_1,v) \in \mathcal{Z}%
_1 \times \mathcal{V}: (x,z_1,v) \in \mathcal{XZ}_1\mathcal{V} \}$.

\begin{assumption}[Common Support]
\label{assumption:cs} $\mathcal{Z}_1\mathcal{V}(x) = \mathcal{Z}_1\mathcal{V}
$.
\end{assumption}
The main implication of common support is the identification of $\theta(x)$
from the identification of $\theta(x,z_1,v)$ in $(z_1,v) \in \mathcal{Z}_1%
\mathcal{V}(x) = \mathcal{Z}_1\mathcal{V}$. 
Assumption \ref{assumption:cs} is only plausible under exclusion
restrictions on $Z_2$ with respect to $X$; see the example in Remark \ref{er}%
. We now establish the identification of the typical global effect \eqref{ge}%
.

\begin{theorem}[Identification of Global Effects]
If $\theta(x,z_1,v)$ is identified for all $(x,z_1,v) \in \mathcal{XZ}_1%
\mathcal{V}$, then $\theta_S(x)$ is identified for all $x \in \mathcal{X}$
that satisfy Assumption \ref{assumption:cs}.
\end{theorem}
We can now apply this result to show the identification of global effects in
the selected population, because under Assumption \ref{assumption:cv}, the
local effects are identified over $\mathcal{XZ}_1\mathcal{V}$, which is the
support of $(X,Z_1,V)$ in the selected population.

\begin{remark}[Global Effects in the Entire Population]
\label{remark::entire_population} The effects in the selected population
generally differ from the effects in the entire population, except under the
additional support condition: 
\begin{equation}
\mathcal{Z}_1\mathcal{V}=\mathcal{Z}_1\mathcal{V}^0,  \label{ass:fs}
\end{equation}%
where $\mathcal{Z}_1\mathcal{V}^0$ is the support of $(Z_1,V)$ in the entire
population. 
This condition requires an excluded variable in $Z_2$ with sufficient
variation to make $h(Z,\eta )>0$ for any $(z_1,\eta)$. 
\end{remark}

\subsection{Global effects for the treated and average derivatives}

\label{ss:global_treated}

Assumption \ref{assumption:cs} might be too restrictive for empirical
applications where an excluded variable with large support is not available.
Without this assumption the global effects are not point identified. 
The alternative generic global effect: 
\begin{equation}
\theta _{S}(x\mid x_{0})=\int \theta (x,z_{1},v)dF_{Z_{1},V\mid
X,C>0}(z_{1},v\mid x_{0}),  \label{ge2}
\end{equation}%
is point identified under weaker support conditions than (\ref{ge}).
Examples of \eqref{ge2} include the ASF conditional on $X=x_{0}$ in the
selected population: 
\begin{equation*}
\mu _{S}(x\mid x_{0})=\mathbb{E}[g(x,Z_{1},\varepsilon )\mid X=x_{0},C>0],
\end{equation*}%
which is a special case of \eqref{ge2} with $\theta (x,z_{1},v)=\mu
(x,z_{1},v)$. This ASF measures the mean of the potential outcome $%
g(x,Z_{1},\varepsilon )$ for the selected individuals with $X=x_{0}$, and is
useful to construct the average treatment effect on the treated of changing $%
X$ from $x_{0}$ to $x_{1}$: 
\begin{equation}
\mu _{S}(x_{1}\mid x_{0})-\mu _{S}(x_{0}\mid x_{0}).  \label{eq:estimated1}
\end{equation}%
%
%
%
%
%
%
%
%
%
%
%
%
%
%
%
%
%
%
%
%
%
%
%
%
%
%
The object in \eqref{ge2} is identified in the selected population under the
following support condition:

\begin{assumption}[Weak Common Support]
\label{assumption:wcs} $\mathcal{Z}_1\mathcal{V}(x) \supseteq \mathcal{Z}_1%
\mathcal{V}(x_0)$.
\end{assumption}
Assumption \ref{assumption:wcs} is weaker than Assumption \ref{assumption:cs}
because $\mathcal{Z}_1\mathcal{V}(x_{0}) \subseteq \mathcal{Z}_1\mathcal{V}$%
. In particular, if the selection equation \eqref{2} is increasing in $X$
and $\mathcal{X}$ is bounded from below, then Assumption \ref{assumption:wcs}
is satisfied by setting $x_0$ lower than $x$.

We define the $\tau$-quantile treatment on the treated as: 
\begin{equation*}
q_{S}(\tau,x_1 \mid x_0)-q_{S}(\tau, x_{0} \mid x_0),
\end{equation*}%
where $q_{S}(\tau, x \mid x_0)$ is the left-inverse of the DSF conditional
on $X=x_0$ in the selected population, 
\begin{equation*}
G_{S}(y,x \mid x_0):=\mathbb{E}[1\{g(x,Z_1,\varepsilon )\leq y\}\mid X=x_0,
C>0],
\end{equation*}%
which is a special case of the effect \eqref{ge2} with $\theta(x,z_1,v) =
G(y,x,z_1,v)$. 

We now establish the identification of the typical global effect \eqref{ge2}.

\begin{theorem}[Identification of Global effects for the Treated]
If $\theta (x,z_1,v)$ is identified for all $(x,z_1,v)\in \mathcal{XZ}_1%
\mathcal{V}$, then $\theta _{S}(x\mid x_{0})$ is identified for all $x\in 
\mathcal{X}$ that satisfy Assumption \ref{assumption:wcs}.
\end{theorem}
We can define global objects in the selected population that are identified
without a common support assumption when $X$ is continuous and $x\mapsto
g(x,\cdot )$ is differentiable. One example is the average derivative
conditional on $X=x$ in the selected population: 
\begin{equation*}
\delta _{S}(x)=\mathbb{E}[\delta (x,Z_1,V)\mid X=x,C>0],
\end{equation*}%
which is a special case of the effect \eqref{ge2} with $\theta
(x,z_1,v)=\delta (x,v)$ and $x_{0}=x$. This object is point identified in
the selected population under Assumption \ref{assumption:cv} because the
integral is over $\mathcal{Z}_1\mathcal{V}(x)$, the support of $(Z_1,V)$
conditional on $X=x$ in the selected population. Another example is the
average derivative in the selected population: 
\begin{equation*}
\delta _{S}=\mathbb{E}[\delta (X,Z_1,V)\mid C>0],
\end{equation*}%
which is point identified under Assumption \ref{assumption:cv} because the
integral is over $\mathcal{XZ}_1\mathcal{V}$, the support of $(X,Z_1,V)$ in
the selected population. This is a special case of the generic global
effect: 
\begin{equation}
\theta _{S}=\int \theta (x,z_1,v)dF_{X,Z_1,V \mid C>0}(x,z_1,v).  \label{ge3}
\end{equation}


\subsection{Bounds for the global effects}

\label{ss:bounds} Another approach to relaxing the support condition
required by Assumption \ref{assumption:cs} is to construct bounds on the
structural functions. As bounds are easily obtained for the LDSF, which is
restricted between 0 and 1, we only discuss the bounds for the DSF noting
that those for the ASF are similar.

We start by separating the identified part of the DSF from the unidentified
part: 
\begin{equation*}
G_{S}(y,x)=\int_{\mathcal{Z}_{1}\mathcal{V}(x)}G(y,x,z_{1},v)dF_{Z_{1},V\mid
C>0}(z_{1},v)+\int_{\overline{\mathcal{Z}_{1}\mathcal{V}}%
(x)}G(y,x,z_{1},v)dF_{Z_{1},V\mid C>0}(z_{1},v)
\end{equation*}%
where $\overline{\mathcal{Z}_{1}\mathcal{V}}(x)=\mathcal{Z}_{1}\mathcal{V}%
\setminus \mathcal{Z}_{1}\mathcal{V}(x).$ Then, following Imbens and Newey
(2009), we form the bounds as: 
\begin{multline*}
\int_{\mathcal{Z}_{1}\mathcal{V}(x)}G(y,x,z_{1},v)dF_{Z_{1},V\mid
C>0}(z_{1},v)\leq G_{S}(y,x)\leq \\
\int_{\mathcal{Z}_{1}\mathcal{V}(x)}G(y,x,z_{1},v)dF_{Z_{1},V\mid
C>0}(z_{1},v)+\int_{\overline{\mathcal{Z}_{1}\mathcal{V}}(x)}dF_{Z_{1},V\mid
C>0}(z_{1},v),
\end{multline*}%
since $0\leq G(y,x,z_{1},v)\leq 1$. This implies that the width of the
bounds equals $P[(Z_{1},V)\in \overline{\mathcal{Z}_{1}\mathcal{V}}(x)\mid
C>0]$. An alternative approach, suggested by Newey (2007) for binary
selection, employs the propensity score $P=P(C>0\mid Z)$ as a control
function. 
In Appendix \ref{app:tighter_bounds}, we show that our bounds are tighter
than the bounds obtained from the propensity score. This highlights another
benefit of using a censored, rather than a binary selection equation, when
possible.

It is also possible to calculate bounds for the DSF for the entire
population since:\label{insert:entire_population} 
\begin{multline*}
G(y,x):=\mathbb{E}[1\{g(x,Z_{1},\varepsilon )\leq y\}]=\int_{\mathcal{Z}_{1}%
\mathcal{V}(x)}G(y,x,z_{1},v)dF_{Z_{1},V}(z_{1},v) \\
+\int_{\mathcal{Z}_{1}\mathcal{V}^{0}\setminus \mathcal{Z}_{1}\mathcal{V}%
(x)}G(y,x,z_{1},v)dF_{Z_{1},V}(z_{1},v),
\end{multline*}%
and hence: 
\begin{multline*}
\int_{\mathcal{Z}_{1}\mathcal{V}(x)}G(y,x,z_{1},v)dF_{Z_{1},V}(z_{1},v)\leq
G(y,x)\leq \int_{\mathcal{Z}_{1}\mathcal{V}%
(x)}G(y,x,z_{1},v)dF_{Z_{1},V}(z_{1},v) \\
+\int_{\mathcal{Z}_{1}\mathcal{V}^{0}\setminus \mathcal{V}%
(x)}dF_{Z_{1},V}(z_{1},v).
\end{multline*}%
The ability to construct bounds for the entire population reflects that the
local effects are not conditional on the selected population.

\section{Estimation and inference}

\label{sec:estimation}

The effects of interest are all identified by functionals of the
distribution of the observed variables and the control function in the
selected population. The control function is the distribution of the
censoring variable $C$ conditional on all explanatory variables $Z$. We
propose a multistep semiparametric method based on least squares,
distribution and quantile regressions to estimate the effects. The reduced
form specifications used in each step can be motivated by parametric
restrictions on the model \eqref{1}--\eqref{2}. We refer to Chernozhukov et
al. (2020) for examples of such restrictions. The semiparametric structure
produces precise estimators and dispenses with the support assumptions for
the identification of the global effects (Newey and Stouli, 2018). In
practice, we recommend to supplement the estimators of the global effects
with estimators of their bounds that do not rely on the semiparametric
structure to avoid the support assumptions. We provide an example of this
robustness check in Section \ref{sec:empirical_example}.


Throughout this section, we assume that we have a random sample of size $n$, 
$\{(Y_{i}\times 1(C_{i}>0),C_{i},Z_{i})\}_{i=1}^{n}$, of the random
variables $(Y\times 1(C>0),C,Z)$, where $Y\times 1(C>0)$ indicates that $Y$
is observed only when $C>0$.


\subsection{Step 1: Estimation of the control function}

\label{ss:step_1}

We estimate the control function using logistic distribution regression
(Foresi and Peracchi, 1995, and Chernozhukov et al., 2013). For every
observation in the selected sample, we set: 
\begin{equation*}
\widehat{V}_{i}=\Lambda (R_{i}^{\mathrm{T}}\widehat{\pi }(C_{i})),\ \
R_{i}:=r(Z_{i}),\ \ i=1,\ldots ,n,C_{i}>0,
\end{equation*}%
where, for $c\in \mathcal{C}_{n},$ the empirical support of $C$, 
\begin{equation*}
\widehat{\pi }(c) \in \arg \max_{\pi \in \mathbb{R}^{d_{r}}}\sum_{i=1}^{n}%
\left[ 1\{C_{i}\leq c\}\log \Lambda (R_{i}^{\mathrm{T}}\pi
))+1\{C_{i}>c\}\log \Lambda (-R_{i}^{\mathrm{T}}\pi )\right] ,
\end{equation*}%
$\Lambda $ is the logistic distribution and $r(z)$ is a $d_{r}$-dimensional
vector of transformations of $z$ with good approximating properties such as
polynomials, B-splines and interactions. 

\subsection{Step 2: Estimation of local objects}

\label{ss:local_derivatives}

We can estimate the local average, distribution and quantile structural
functions using flexibly parametrized least squares, distribution and
quantile regressions, where we replace the control function by its estimator
from the previous step. %
For technical reasons explained in Section \ref{ss:asymptotic_theory}, our
estimation method is based on a trimmed sample with respect to the censoring
variable $C $. Therefore, we introduce the following trimming indicator
among the selected sample: 
\begin{equation*}
T=1(C\in \overline{\mathcal{C}})
\end{equation*}%
where $\overline{\mathcal{C}}=(0,\overline{c}]$ for some $0<\overline{c}%
<\infty $, such that $P(T=1)>0$.

The estimator of the LASF is $\widehat{\mu }(x,z_{1},v)=w(x,z_{1},v)^{%
\mathrm{T}}\widehat{\beta },$ where $w(x,z_{1},v)$ is a $d_{w}$-dimensional
vector of transformations of $(x,z_{1},v)$ with good approximating
properties, and $\widehat{\beta }$ is the ordinary least squares estimator:%
\footnote{%
An alternative approach is to follow Jun (2009) and Masten and Torgovitsky
(2016). These papers acknowledge that with an index restriction the
parameters of interest can be estimated in the presence of a control
function by estimation over subsamples for which the control function has a 
\emph{similar} value. While each of these papers considers a random
coefficients model with endogeneity their approach is applicable here.} 
\begin{equation*}
\widehat{\beta }=\left[ \sum_{i=1}^{n}\widehat{W}_{i}\widehat{W}_{i}^{%
\mathrm{T}}T_{i}\right] ^{-1}\sum_{i=1}^{n}\widehat{W}_{i}Y_{i}T_{i},\ \ 
\widehat{W}_{i}:=w(X_{i},Z_{1i},\widehat{V}_{i}),
\end{equation*}%
provided that $\sum_{i=1}^{n}\widehat{W}_{i}\widehat{W}_{i}^{\mathrm{T}%
}T_{i} $ is invertible. The estimator of the LDSF is $\widehat{G}%
(y,x,z_{1},v)=\Lambda (w(x,z_{1},v)^{\mathrm{T}}\widehat{\beta }(y))$, where 
$\widehat{\beta }(y)$ is the logistic distribution regression estimator: 
\begin{equation*}
\widehat{\beta }(y)\in \arg \max_{b\in \mathbb{R}^{d_{w}}}\sum_{i=1}^{n}%
\left[ 1\{Y_{i}\leq y\}\log \Lambda (\widehat{W}_{i}^{\mathrm{T}%
}b))+1\{Y_{i}>y\}\log \Lambda (-\widehat{W}_{i}^{\mathrm{T}}b))\right] T_{i}.
\end{equation*}%
Similarly, the estimator of the LQSF is $\widehat{q}(\tau
,x,z_{1},v)=w(x,z_{1},v)^{\mathrm{T}}\widehat{\beta }(\tau )$, where $%
\widehat{\beta }(\tau )$ is the Koenker and Bassett (1978) quantile
regression estimator: 
\begin{equation*}
\widehat{\beta }(\tau )\in \arg \min_{b\in \mathbb{R}^{d_{w}}}\sum_{i=1}^{n}%
\rho _{\tau }(Y_{i}-\widehat{W}_{i}^{\mathrm{T}}b)T_{i}.
\end{equation*}%
Estimators of the local derivatives are obtained by taking derivatives of
the estimators of the local structural functions. Thus, the estimator of the
LADF is: 
\begin{equation*}
\widehat{\delta }(x,z_{1},v)=\partial _{x}w(x,z_{1},v)^{\mathrm{T}}\widehat{%
\beta },
\end{equation*}%
and the estimator of the LQDF is: 
\begin{equation*}
\widehat{\delta }_{\tau }(x,z_{1},v)=\partial _{x}w(x,z_{1},v)^{\mathrm{T}}%
\widehat{\beta }(\tau ).
\end{equation*}


\subsection{Step 3: Estimation of global effects}

We obtain estimators of the generic global effects by approximating the
integrals over the control function by averages of the estimated local
effects evaluated at the estimated control function. The estimator of the
effect \eqref{ge} is: 
\begin{equation*}
\widehat{\theta }_{S}(x)=\sum_{i=1}^{n}T_{i}\widehat{\theta }(x,Z_{1i},%
\widehat{V}_{i})/\sum_{i=1}^{n}T_{i}.
\end{equation*}%
This yields the estimators of the ASF for $\widehat{\theta }(x,z_1,v)=%
\widehat{\mu }(x,z_1,v)$ and DSF at $y$ for $\widehat{\theta }(x,z_1,v)=%
\widehat{G}(y,x,z_1,v)$. The estimator of the QSF is then obtained by
inversion of the estimator of the DSF.\footnote{%
We use the generalized inverse 
\begin{equation*}
\widehat{q}_{S}(\tau ,x)=\int_{0}^{\infty }1(\widehat{G}_{S}(y,x)\leq \tau
)dy-\int_{-\infty }^{0}1(\widehat{G}_{S}(y,x)>\tau )dy,
\end{equation*}%
which does not require that the estimator of the DSF $y\mapsto \widehat{G}%
_{S}(y,x)$ be monotone.} We form an estimator of the effect \eqref{ge2} as: 
\begin{equation*}
\widehat{\theta }_{S}(x\mid x_{0})=\sum_{i=1}^{n}T_{i}K_{i}(x_{0})\widehat{%
\theta }(x,Z_{1i},\widehat{V}_{i})/\sum_{i=1}^{n}T_{i}K_{i}(x_{0}),
\end{equation*}%
for $K_{i}(x_{0})=1(X_{i}=x_{0})$ when $X$ is discrete or $%
K_{i}(x_{0})=k_{h}(X_{i}-x_{0})$ when $X$ is continuous, where $%
k_{h}(u)=k(u/h)/h$, $k$ is a kernel, and $h$ is a bandwidth such as $%
h\rightarrow 0$ as $n\rightarrow \infty$. Finally, the estimator of the
effect \eqref{ge3} is: 
\begin{equation*}
\widehat{\theta }_{S}=\sum_{i=1}^{n}T_{i}\widehat{\theta }(X_{i},Z_{1i},%
\widehat{V}_{i})/\sum_{i=1}^{n}T_{i}.
\end{equation*}

\subsection{Estimation of bounds on global effects}

We provide estimators of the bounds of $G(y,x)$. Estimators of the bounds of 
$G_S(y,x)$ can be constructed similarly. We start by rewriting the width of
the bounds as 
\begin{equation*}
\int_{\mathcal{Z}_{1}\mathcal{V}^{0}\setminus \mathcal{V}%
(x)}dF_{Z_{1},V}(z_{1},v)= \int_{\mathcal{Z}_{1}\mathcal{V}^{0}} \mathbf{1}%
(V > v_u(z_1,x), V < v_{\ell}(z_1,x) ) dF_{Z_{1},V}(z_{1},v),
\end{equation*}
where $v_u(x,z_1) = \sup\{v : (x,z_1,v)\in \mathcal{XZ}_1\mathcal{V}\}$ and $%
v_{\ell}(x,z_1) = \inf\{v : (x,z_1,v)\in \mathcal{XZ}_1\mathcal{V}\}$. By
the monotonicity in Assumption \ref{assumption:cv}, $v_u(x,z_1) = 1$ and $%
v_{\ell}(x,z_1) = \inf\{ F_{Z}(0 \mid Z=z) : z \in \mathcal{Z}%
_{2}(x,z_{1})\} $, where $\mathcal{Z}_{2}(x,z_{1})$ is the support of $Z_{2}
\mid X=x,Z_{1}=z_{1}$. We can then estimate $v_{\ell}(x,z_1)$ by 
\begin{equation*}
\widehat v_{\ell}(x,z_1) = \min \{ \Lambda (r(Z_i)^{\mathrm{T}}\widehat{\pi }%
(0)) : Z_{1i} = z_1, Z_{2i} \supseteq x, i = 1,\ldots, n \}.
\end{equation*}

Finally, we can form estimators of the lower and upper bounds of $G(y,x)$
as: 
\begin{equation*}
\frac{1}{n} \sum_{i=1}^{n}\mathbf{1}(\widehat V_{i}\geq \widehat
v_{\ell}(x,Z_{1i}))\widehat{G}(y,x,Z_{1i}, V_i)
\end{equation*}
and 
\begin{equation*}
\frac{1}{n} \sum_{i=1}^{n}\mathbf{1}(\widehat V_{i}\geq \widehat
v_{\ell}(x,Z_{1i}))\widehat{G}(y,x,Z_{1i},V_i)+\frac{1}{n} \sum_{i=1}^{n}%
\mathbf{1}(\widehat V_{i}<\widehat v_{\ell}(x,Z_{1i})).
\end{equation*}


\subsection{Inference}

We use weighted bootstrap to make inference on all objects of interest
(Praestgaard and Wellner, 1993; Hahn, 1995). This method obtains the
bootstrap version of the estimator of interest by repeating all estimation
steps including random draws from a distribution as sampling weights. The
weights should be positive and come from a distribution with unit mean and
variance such as the standard exponential. The weighted bootstrap has some
theoretical and practical advantages over the empirical bootstrap. Thus, it
is appealing that the consistency can be proven following the strategy set
forth by Ma and Kosorok (2005) and the smoothness induced by the weights
helps dealing with discrete covariates with small cell sizes. The
implementation of the bootstrap for the local and global effects is
summarized in the following algorithm:

\begin{algorithm}[Weighted Bootstrap]
For $b=1,\ldots ,B$, repeat the following steps: (1) Draw a set of weights $%
(\omega _{1}^{b},\ldots ,\omega _{n}^{b})$ i.i.d. from a distribution that
satisfies Condition \ref{ass:sampling}(b), stated in the following section,
such as the standard exponential distribution. (2) Obtain the bootstrap
draws of the control function, $\widehat{V}_{i}^{b}=\Lambda (R_{i}^{\mathrm{T%
}}\widehat{\pi }^{b}(C_{i}))$, $i=1,\ldots ,n$, where for $c\in \mathcal{C}%
_{n},$ 
\begin{equation*}
\widehat{\pi }^{b}(c)=\arg \max_{\pi \in \mathbb{R}^{d_{r}}}\sum_{i=1}^{n}%
\omega _{i}^{b}\left[ 1\{C_{i}\leq c\}\log \Lambda (R_{i}^{\mathrm{T}}\pi
))+1\{C_{i}>c\}\log \Lambda (-R_{i}^{\mathrm{T}}\pi )\right] .
\end{equation*}%
(3) Obtain the bootstrap draw of the local effect, $\widehat{\theta }%
^{b}(x,z_1,v)$. For the LASF, $\widehat{\theta }^{b}(x,z_1,v)=\widehat{\mu }%
^{b}(x,z_1,v)=w(x,z_1,v)^{\mathrm{T}}\widehat{\beta }^{b},$ where 
\begin{equation*}
\widehat{\beta }^{b}=\left[ \sum_{i=1}^{n}\omega _{i}^{b}\widehat{W}_{i}^{b}(%
\widehat{W}_{i}^{b})^{\mathrm{T}}T_{i}\right] ^{-1}\sum_{i=1}^{n}\omega
_{i}^{b}\widehat{W}_{i}^{b}Y_{i}T_{i},\ \ \widehat{W}%
_{i}^{b}:=w(X_{i},Z_{1i}, \widehat{V}_{i}^{b}).
\end{equation*}%
For the LDSF, $\widehat{\theta }^{b}(x,z_1,v)=\widehat{G}^{b}(y,x,z_1,v)=%
\Lambda (w(x,z_1,v)^{\mathrm{T}}\widehat{\beta }^{b}(y))$, where 
\begin{equation*}
\widehat{\beta }^{b}(y)=\arg \max_{b\in \mathbb{R}^{d_{w}}}\sum_{i=1}^{n}%
\omega _{i}^{b}\left[ 1\{Y_{i}\leq y\}\log \Lambda (b^{\mathrm{T}}\widehat{W}%
_{i}^{b})+1\{Y_{i}>y\}\log \Lambda (-b^{\mathrm{T}}\widehat{W}_{i}^{b})%
\right] T_{i}.
\end{equation*}
For the LQSF, $\widehat{\theta }^{b}(x,z_1,v)=\widehat{q}^{b}(\tau
,x,z_1,v)=w(x,z_1,v)^{\mathrm{T}}\widehat{\beta }^{b}(\tau )$, where 
\begin{equation*}
\widehat{\beta }^{b}(\tau )=\arg \min_{b\in \mathbb{R}^{d_{w}}}%
\sum_{i=1}^{n}\omega _{i}^{b}\rho _{\tau }(Y_{i}-b^{\mathrm{T}}\widehat{W}%
_{i}^{b})T_{i}.
\end{equation*}%
(4) Obtain the bootstrap draw of the global effects as 
\begin{equation*}
\widehat{\theta }_{S}^{b}(x)=\sum_{i=1}^{n}\omega _{i}^{b}T_{i}\widehat{%
\theta }^{b}(x,Z_{1i},\widehat{V}_{i}^{b})/\sum_{i=1}^{n}\omega
_{i}^{b}T_{i},
\end{equation*}%
\begin{equation*}
\widehat{\theta }_{S}^{b}(x\mid x_{0})=\sum_{i=1}^{n}\omega
_{i}^{b}T_{i}K_{i}(x_{0})\widehat{\theta }^{b}(x,Z_{1i},\widehat{V}%
_{i}^{b})/\sum_{i=1}^{n}\omega _{i}^{b}T_{i}K_{i}(x_{0}),
\end{equation*}%
or 
\begin{equation*}
\widehat{\theta }_{S}^{b}=\sum_{i=1}^{n}\omega _{i}^{b}T_{i}\widehat{\theta }%
^{b}(X_{i},Z_{1i},\widehat{V}_{i}^{b})/\sum_{i=1}^{n}\omega _{i}^{b}T_{i}.
\end{equation*}
\end{algorithm}

\subsection{Asymptotic theory}

\label{ss:asymptotic_theory}

We derive large sample theory for the local and global effects. We focus on
the average effects for the sake of brevity. The theory for distribution and
quantile effects can be derived using similar arguments; see, for example,
Chernozhukov et al. (2015) and Chernozhukov et al. (2020).\footnote{%
Chernozhukov et al. (2015) and Chernozhukov et al. (2020) considered models
with endogeneity and censoring, and models with endogeneity, respectively.
They derived theory for two-step estimators of these models where the first
step is estimated by distribution regression and the second step by quantile
or distribution regression. We derive theory for two-step estimators of
sample selection models where the first step is estimated by distribution
regression and the second step by least squares.} Throughout the analysis,
we assume that the flexible specifications used in all the steps are correct
and treat their dimensions as fixed, so that all model parameters are
estimable at a $\sqrt{n} $-rate and the estimation of the global effects
does not require any support assumptions.


In what follows we use the following notation. We let the random vector $%
A=(Y\times 1(C>0),C,Z,V)$ live on some probability space $(\Omega _{0},%
\mathcal{F}_{0},P)$. Thus, the probability measure $P$ determines the law of 
$A$ or any of its elements. We also let $A_{1},...,A_{n}$ be i.i.d. copies
of $A$ and live on the complete probability space $(\Omega ,\mathcal{F},\Pr
) $, which contains the infinite product of $(\Omega _{0},\mathcal{F}_{0},P)$%
. Moreover, this probability space can be suitably enriched to also carry
the random weights that appear in the weighted bootstrap. The distinction
between the two laws $P$ and $\Pr $ is helpful to simplify the notation in
the proofs and in the analysis. Calligraphic letters such as $\mathcal{Y}$
and $\mathcal{X}$ denote the supports of $Y\times 1(C>0)$ and $X$; and $%
\mathcal{YX}$ denotes the joint support of $(Y,X)$. 
Unless explicitly mentioned, all functions appearing in the statements are
assumed to be measurable.

We now formally state the assumptions. The first assumption is about
sampling and the bootstrap weights.

\begin{condition}[Sampling and Bootstrap Weights]
\label{ass:sampling} (a) Sampling: the data $\{Y_{i}\times
1(C_{i}>0),C_{i},Z_{i}\}_{i=1}^{n}$ are a sample of size $n$ of independent
and identically distributed observations from the random vector $(Y\times
1(C>0),C,Z).$ (b) Bootstrap weights: $(\omega_{1},...,\omega_{n})$ are
i.i.d. draws from a random variable $\omega \geq 0$, with ${\mathbb{E}}%
_{P}[\omega]=1$, $\mathrm{Var}_{P}[\omega]=1,$ and ${\mathbb{E}}%
_{P}|\omega|^{2+\delta }<\infty $ for some $\delta >0 $; live on the
probability space $(\Omega ,\mathcal{F},\Pr )$; and are independent of the
data $\{Y_{i}\times 1(C_{i}>0),C_{i},Z_{i}\}_{i=1}^{n}$ for all $n$.
\end{condition}
The second assumption concerns the first stage in which we estimate the
control function: 
\begin{equation*}
\vartheta _{0}(c,z):=F_{C}(c\mid z).
\end{equation*}%
We assume a logistic distribution regression model for the conditional
distribution of $C$ in the trimmed support, $\overline{\mathcal{C}}$, that
excludes censored and extreme values of $C$. The purpose of the upper
trimming is to avoid the upper tail in the modeling and estimation of the
control function, and to make the eigenvalue assumption in Condition \ref%
{ass:first}(b) more plausible. We consider a fixed trimming rule, which
greatly simplifies the derivation of the asymptotic properties.\footnote{%
We view the fixed trimming as a convenient theoretical device that is not
restrictive in practical applications. Indeed, our estimators perform well
without trimming in numerical simulations. Alternative random, data driven
rules are possible at the cost of more complicated proofs and technical
conditions on the upper tail of $F_{C}(c\mid z)$.} Throughout this section,
we use bars to denote trimmed supports with respect to $C$, e.g., $\overline{%
\mathcal{C}\mathcal{Z}}=\{(c,z)\in \mathcal{C}\mathcal{Z}:c\in \overline{%
\mathcal{C}}\}$, and $\overline{\mathcal{V}}=\{\vartheta _{0}(c,z):(c,z)\in 
\overline{\mathcal{C}\mathcal{Z}}\}$.

\begin{condition}[First Stage]
\label{ass:first} (a) Trimming: we consider the trimming rule as defined by
the indicator $T=\mathbf{1}(C \in \overline{ \mathcal{C}})$. 
(b) Model: the distribution of $C$ conditional on $Z$ follows the
distribution regression model in the trimmed support $\overline{\mathcal{C}}$%
, \emph{i.e.}, 
\begin{equation*}
F_C(c \mid Z) = F_C(c \mid R) = \Lambda(R^{\mathrm{T}}\pi_0(c)), \ \ R =
r(Z),
\end{equation*}
for all $c \in \overline{\mathcal{C}}$, where $\Lambda$ is the logit link
function; the coefficients $c \mapsto \pi_0(c)$ are three times continuously
differentiable with uniformly bounded derivatives; $\overline{\mathcal{R}}$
is compact; and the minimum eigenvalue of ${\mathbb{E}}_P \left[\Lambda(R^{%
\mathrm{T}}\pi_0(c))[1- \Lambda(R^{\mathrm{T}}\pi_0(c))] RR^{\mathrm{T}} %
\right] $ is bounded away from zero uniformly over $c \in \overline{\mathcal{%
C}}$.
\end{condition}

For $c\in \overline{\mathcal{C}}$, let: 
\begin{equation*}
\widehat{\pi }^{b}(c)\in \arg \min_{\pi \in \mathbb{R}^{\dim
(R)}}\sum_{i=1}^{n}\omega _{i}\{1(C_{i}\leq c)\log \Lambda (R_{i}^{\mathrm{T}%
}\pi )+1(C_{i}>c)\log \Lambda (-R_{i}^{\mathrm{T}}\pi )\},
\end{equation*}%
where either $\omega _{i}=1$ for the unweighted sample, to obtain the
estimator, or $\omega _{i}$ are the bootstrap weights for obtaining
bootstrap draws of the estimator. Then set: 
\begin{equation*}
\vartheta _{0}(c,r)=\Lambda (r^{\mathrm{T}}\pi _{0}(c));\ \widehat{\vartheta 
}^{b}(c,r)=\Lambda (r^{\mathrm{T}}\widehat{\pi }^{b}(c)),
\end{equation*}%
if $(c,r)\in \overline{\mathcal{C}\mathcal{R}},$ and $\vartheta _{0}(c,r)=%
\widehat{\vartheta }^{b}(c,r)=0$ otherwise.

Theorem 4 of Chernozhukov et al. (2015) established the asymptotic
properties of the DR estimator of the control function. We repeat the result
here as a lemma for completeness and to introduce notation that will be used
in the results below. Let $\|f\|_{T,\infty} := \sup_{a \in \mathcal{A}}
|T(c) f(a)|$ for any function $f : \mathcal{A} \mapsto \mathbb{R}$, $%
\ell^{\infty}(\mathcal{A})$ be the set of bounded functions on $\mathcal{A}$
equipped with the norm $\| \cdot \|_{T,\infty}$, and $\lambda =
\Lambda(1-\Lambda)$, the density of the logistic distribution.

\begin{lemma}[First Stage]
\label{lemma:first} Suppose that Conditions \ref{ass:sampling} and \ref%
{ass:first} hold. Then, (1) 
\begin{eqnarray*}
\sqrt{n}(\widehat{\vartheta }^{b}(c,r)-\vartheta _{0}(c,r)) &=&\frac{1}{%
\sqrt{n}}\sum_{i=1}^{n}e_{i}\ell (A_{i},c,r)+o_{\Pr }(1)\rightsquigarrow
\Delta ^{b}(c,r)\text{ in }\ell ^{\infty }(\overline{\mathcal{C}\mathcal{R}}%
), \\
\ell (A,c,r):= &&\lambda (r^{\mathrm{T}}\pi _{0}(c))[1\{C\leq c\}-\Lambda
(R^{\mathrm{T}}\pi _{0}(c))]\times \\
&&\times r^{\mathrm{T}}{\mathbb{E}}_{P}\left\{ \Lambda (R^{\mathrm{T}}\pi
_{0}(c))[1-\Lambda (R^{\mathrm{T}}\pi _{0}(c))]RR^{\mathrm{T}}\right\}
^{-1}R, \\
{\mathbb{E}}_{P}[\ell (A,c,r)] &=&0,{\mathbb{E}}_{P}[T\ell
(A,C,R)^{2}]<\infty ,
\end{eqnarray*}%
where $(c,r)\mapsto \Delta ^{b}(c,r)$ is a Gaussian process with uniformly
continuous sample paths and covariance function given by ${\mathbb{E}}%
_{P}[\ell (A,c,r)\ell (A,\tilde{c},\tilde{r})^{\mathrm{T}}]$. 
(2) There exists $\widetilde{\vartheta }^{b}:\overline{\mathcal{CR}}\mapsto
\lbrack 0,1]$ that obeys the same first-order representation uniformly over $%
\overline{\mathcal{CR}}$, is close to $\widehat{\vartheta }^{b}$ in the
sense that $\Vert \widetilde{\vartheta }^{b}-\widehat{\vartheta }^{b}\Vert
_{T,\infty }=o_{\Pr }(1/\sqrt{n})$ and, with probability approaching one,
belongs to a bounded function class $\Upsilon $ such that the covering
entropy satisfies:\footnote{%
See Appendix \ref{app:estimation} for a definition of the covering entropy.} 
\begin{equation*}
\log N(\epsilon ,\Upsilon ,\Vert \cdot \Vert _{T,\infty })\lesssim \epsilon
^{-1/2},\ \ 0<\epsilon <1.
\end{equation*}
\end{lemma}
The next assumptions are about the second stage. We assume a flexible linear
model for the conditional distribution of $Y$ given $(X,V)$ in the trimmed
support $C \in \overline{\mathcal{C}}$, impose compactness conditions, and
provide sufficient conditions for the identification of the parameters.
Compactness is imposed over the trimmed support and can be relaxed at the
cost of more complicated and cumbersome proofs.

\begin{condition}[Second Stage]
\label{ass:second} (a) Model: the expectation of $Y$ conditional on $(X,Z_1,$
$V)$ in the trimmed support $C\in \overline{\mathcal{C}}$ is: 
\begin{equation*}
\mathbb{E}(Y\mid X,Z_1,V,C\in \overline{\mathcal{C}})=W^{\mathrm{T}}\beta
_{0},\ \ V=F_{C|Z}(C\mid Z),\ \ W=w(X,Z_1,V).
\end{equation*}%
(b) Compactness and moments: the set $\overline{\mathcal{W}}$ is compact; 
the derivative vector $\partial _{v}w(x,z_1,v)$ exists and its components
are uniformly continuous in $v\in \overline{\mathcal{V}}$, uniformly in $%
(x,z_1)\in \overline{\mathcal{XZ}}_1$, and are bounded in absolute value by
a constant, uniformly in $(x,z_1,v)\in \overline{\mathcal{XZ}_1\mathcal{V}}$%
; $\mathbb{E}(Y^{2}\mid C\in \overline{\mathcal{C}})<\infty $; and $\beta
_{0}\in \mathcal{B}$, where $\mathcal{B}$ is a compact subset of $\mathbb{R}%
^{d_{w}}$. (c) Identification and nondegeneracy: the matrix $J:={\mathbb{E}}%
_{P}[WW^{\mathrm{T}}\ T]$ is of full rank; and the matrix $\Omega :=\mathrm{%
Var}_{P}[f_{1}(A)+f_{2}(A)]$ is finite and is of full rank, where: 
\begin{equation*}
f_{1}(A):=\{W^{\mathrm{T}}\beta _{0}-Y\}WT,
\end{equation*}%
and, for $\dot{W}=\partial _{v}w(X,Z_1,v)|_{v=V}$, 
\begin{equation*}
f_{2}(A):={\mathbb{E}}_{P}[\{[W^{\mathrm{T}}\beta _{0}-Y]\dot{W}+W^{\mathrm{T%
}}\beta _{0}W\}T\ell (a,C,R)]\big|_{a=A}.
\end{equation*}
\end{condition}

Let 
\begin{equation*}
\widehat{\beta } \in \arg \min_{\beta \in \mathbb{R}^{\dim(W)}}
\sum_{i=1}^{n} T_i (Y_{i} - \beta^{\mathrm{T}}\widehat W_{i})^2, \ \ 
\widehat{W}_{ i} = w(X_i, Z_{1i}, \widehat V_{ i}), \ \ \widehat V_{i} =
\widehat \vartheta(C_i,R_i),
\end{equation*}
where $\widehat \vartheta$ is the estimator of the control function in the
unweighted sample; and 
\begin{equation*}
\widehat{\beta }^{b} \in \arg \min_{\beta \in \mathbb{R}^{\dim(W)}}
\sum_{i=1}^{n} \omega_i T_i (Y_{i} -\beta^{\mathrm{T}}\widehat W_{i}^{b
})^2, \ \ \widehat{W}_{ i}^b = w(X_i, Z_{1i}, \widehat V_{ i}^b), \ \
\widehat V_{ i}^b = \widehat \vartheta^b(C_i,R_i),
\end{equation*}
where $\widehat \vartheta^b$ is the estimator of the control function in the
weighted sample. The following lemma establishes a central limit theorem and
a central limit theorem for the bootstrap for the estimator of the
coefficients in the second stage. Let $\rightsquigarrow_{\Pr}$ denote the
bootstrap consistency, \emph{i.e.} weak convergence conditional on the data
in probability as defined in Appendix \ref{app:notation}.

\begin{lemma}[CLT and Bootstrap FCLT for $\protect\widehat{\protect\beta }$]

\label{thm:fclt} Under Conditions \ref{ass:sampling}--\ref{ass:second}, in $%
\mathbb{R}^{d_w}$, 
\begin{equation*}
\sqrt{n}(\widehat\beta - \beta_0) \rightsquigarrow J^{-1} G, \text{\ \ and \
\ } \sqrt{n}(\widehat\beta^{b} - \widehat \beta) \rightsquigarrow_{\Pr}
J^{-1} G,
\end{equation*}
where $G \sim N(0,\Omega)$ and $J$ and $\Omega$ are defined in Assumption %
\ref{ass:second}(c).
\end{lemma}
The properties of the estimator of the LASF, $\widehat \mu(x,z_1,v) =
w(x,z_1,v)^{\mathrm{T}} \widehat \beta$, and its bootstrap version, $%
\widehat \mu^b(x,z_1,v) = w(x,z_1,v)^{\mathrm{T}} \widehat \beta^b$,
constitute a corollary of Lemma \ref{thm:fclt}.

\begin{corollary}[FCLT and Bootstrap FCLT for LASF]
\label{fclt:lasf} Under Assumptions \ref{ass:sampling}--\ref{ass:second}, in 
$\ell (\overline{\mathcal{XZ}_1\mathcal{V}})$, 
\begin{equation*}
\sqrt{n}(\widehat{\mu }(x,z_1,v)-\mu (x,z_1,v))\rightsquigarrow Z(x,z_1,v)%
\text{ and }\sqrt{n}(\widehat{\mu }^{b}(x,z_1,v)-\widehat{\mu }%
(x,z_1,v))\rightsquigarrow _{\Pr }Z(x,z_1,v),
\end{equation*}%
where $(x,z_1,v)\mapsto Z(x,z_1,v):=w(x,z_1,v)^{\mathrm{T}}J^{-1}G$ is a
zero-mean Gaussian process with covariance function: 
\begin{equation*}
\mathrm{Cov}%
_{P}[Z(x_{0},z_{10},v_{0}),Z(x_{1},z_{11},v_{1})]=w(x_{0},z_{10},v_{0})^{%
\mathrm{T}}J^{-1}\Omega J^{-1}w(x_{1},z_{11},v_{1}).
\end{equation*}
\end{corollary}
To obtain the properties of the estimator of the ASFs, we define $%
W_{x}:=w(x,Z_1,V)$, $\widehat{W}_{x}:=w(x, Z_1,\widehat{V})$, and $\widehat{W%
}_{x}^{b}:=w(x,Z_1,\widehat{V}^{b})$. The estimator and its bootstrap draw
of the ASF in the trimmed support, $\mu_S(x)={\ \mathbb{E}}_{P}\{\beta_{0}^{%
\mathrm{T}} W_{x} \mid T = 1\}$, are $\widehat{\mu}_S(x)=\sum_{i=1}^{n}%
\widehat{\beta}^{\mathrm{T}} \widehat{W}_{xi} T_{i}/n_T$, and $\widehat{\mu}%
_S^{b}(x)=\sum_{i=1}^{n}e_{i} \widehat{\beta}^{b T }\widehat{W}_{xi}^{b}
T_{i}/n_T^b$, where $n_T = \sum_{i=1}^n T_i$ and $n^b_T = \sum_{i=1}^n e_i
T_i$. The estimator and its bootstrap draw of the ASF on the treated in the
trimmed support, $\mu_S(x \mid x_0)={\ \mathbb{E}}_{P}\{\beta_{0}^{\mathrm{T}%
} W_{x} \mid T = 1, X = x_0\}$, are $\widehat{\mu}_S(x \mid
x_0)=\sum_{i=1}^{n}\widehat{\beta}^{\mathrm{T}} \widehat{W}_{xi} K_i(x_0)
T_{i}/n_T(x_0)$, and $\widehat{\mu}_S^{b}(x) = \sum_{i=1}^{n}e_{i} \widehat{%
\beta}^{b T}\widehat{W}_{xi}^{b}$ $K_i(x_0)T_{i} / n_T^b(x_0)$, where $%
n_T(x_0) = \sum_{i=1}^n K_i(x_0) T_i $ and $n^b_T(x_0) = \sum_{i=1}^n e_i
K_i(x_0) T_i$. Let $p_T := P(T=1)$ and $p_T(x) := P(T=1, X=x)$. The next
result gives the large sample theory for these estimators. The theory for
the ASF on the treated is derived for $X$ discrete, which is the relevant
case in our empirical application.

\begin{theorem}[FCLT and Bootstrap FCLT for ASF]
\label{fclt:sdf} Under Assumptions \ref{ass:sampling}--\ref{ass:second}, in $%
\ell (\overline{\mathcal{X}})$, 
\begin{equation*}
\sqrt{np_{T}}(\widehat{\mu }_{S}(x)-\mu _{S}(x))\rightsquigarrow Z(x)\text{
and }\sqrt{np_{T}}(\widehat{\mu }_{S}^{b}(x)-\widehat{\mu }%
_{S}(x))\rightsquigarrow _{\Pr }Z(x),
\end{equation*}%
where $x\mapsto Z(x)$ is a zero-mean Gaussian process with covariance
function: 
\begin{equation*}
\mathrm{Cov}_{P}[Z(x_{0}),Z(x_{1})]=\mathrm{Cov}_{P}[W_{x_{0}}^{\mathrm{T}%
}\beta _{0}+\sigma _{x_{0}}(A),W_{x_{1}}^{\mathrm{T}}\beta _{0}(v)+\sigma
_{x_{1}}(A)\mid T=1],
\end{equation*}%
with: 
\begin{equation*}
\sigma _{x}(A)={\mathbb{E}}_{P}\{W_{x}^{\mathrm{T}}T%
\}J^{-1}[f_{1}(A)+f_{2}(A)]+{\mathbb{E}}_{P}\{\dot{W}_{x}^{\mathrm{T}}\beta
_{0}T\ell (a,C,R)\}\big|_{a=A}.
\end{equation*}%
Also, if $p_{T}(x_{0})>0$, in $\ell (\overline{\mathcal{X}})$, 
\begin{equation*}
\begin{split}
\sqrt{np_{T}(x_{0})}(\widehat{\mu }_{S}(x& \mid x_{0})-\mu _{S}(x\mid
x_{0}))\rightsquigarrow Z(x\mid x_{0})\text{ and } \\
\sqrt{np_{T}(x_{0})}(\widehat{\mu }_{S}^{b}(x& \mid x_{0})-\widehat{\mu }%
_{S}(x\mid x_{0}))\rightsquigarrow _{\Pr }Z(x\mid x_{0}),
\end{split}%
\end{equation*}%
where $x\mapsto Z(x\mid x_{0})$ is a zero-mean Gaussian process with
covariance function: 
\begin{equation*}
\mathrm{Cov}_{P}[Z(x_1\mid x_{0}),Z(x_2\mid x_{0})]=\mathrm{Cov}%
_{P}[W_{x_1}^{\mathrm{T}}\beta _{0}+\sigma_{x_1}(A),W_{x_2}^{\mathrm{T}%
}\beta _{0}+\sigma _{x_2}(A)\mid T=1,X=x_{0}],
\end{equation*}%
\end{theorem}
Theorem \ref{fclt:sdf} can be used to construct confidence bands for the
ASFs, $x \mapsto \mu_S(x)$ and $x \mapsto \mu_S(x \mid x_0)$, over regions
of values of $x$ via Kolmogorov-Smirnov type statistics and weighted
bootstrap, and to construct confidence intervals for average treatment
effects, $\mu(x_1) - \mu(x_0)$ and $\mu(x_1 \mid x_0) - \mu(x_0 \mid x_0)$,
via t-statistics and weighted bootstrap.

\section{Application: United Kingdom wage regressions}

\label{sec:empirical_example}

We investigate the impact of human capital on the wage level of female
workers in the United Kingdom (UK). We use data from the Family Expenditure
Survey (FES) for the years 1978 to 1999. Blundell et al. (2003) study male
wage growth and Blundell et al. (2007) examine wage inequality for both
males and females using the same data source. We employ the same data
selection rules and refer the reader to these earlier papers for details.
The FES is a repeated cross section of households and contains detailed
information on the number of weekly hours worked and the hourly wage of the
individual. We restrict the data to those who report an education level and
only include working females who report working weekly hours of 70 or less
and an hourly wage of at least 0.01 pounds. This reduces the total number of
observations from 96,402 to 94,985, corresponding to over 4,100 individuals
per year with an average of 2,600 working.

The outcome variable, $Y$, is the log-hourly wage defined as the nominal
weekly earnings divided by the number of hours worked and deflated by the
quarterly UK retail price index. The number of hours worked, $C$, is defined
as the usual number of hours worked per week{.}\footnote{\label%
{footnote:censoring}Some workers who work very few hours per year may report
zero hours as their usual number of hours per week worked. We do not account
for this source of measurement error.} Following Blundell et al. (2003) we
use the simulated out-of-work benefits income as an exclusion restriction in
the hours equation. We refer to their paper for details and note that the UK
benefits system makes this restriction appropriate as, in contrast to other
European countries, unemployment benefits are not related to income prior to
the period out of work. Blundell et al. (2007) argue in an application
studying the behavior of both males and females that the system of housing
benefits may have a positive relationship with the in-work potential.
However, we do not address this possibility here, although we note that
Blundell et al. (2007) suggest a potential solution using a monotonicity
restriction rather than an exclusion restriction in the hours equation.%
\footnote{%
Given this concern regarding the validity of the exclusion restriction we
exploit the ability to identify the effects of interest via the variability
in hours in the absence of an exclusion restriction. Accordingly we
re-estimated the model with the simulated out-of-work benefits included and
excluded from both steps. Although we do not report them here, the
differences in results across these two identification schemes and those
reported in the paper are minor.\label{insert::footnote_exclusion}
\par
{}} Equations (\ref{1}) and (\ref{2}) characterize the model of Blundell et
al. (2003) when $g$ and $h$ are linear and separable and $\varepsilon $ and $%
\eta $ are normally distributed.

Figure \ref{fig:descriptive}A reports the female participation rate over the
sample period. Figure \ref{fig:descriptive}B reports the average number of
hours for all females and those reporting positive hours, respectively.
Recall that the control function exploits the variation in both the
extensive and the intensive margins of the hours decisions. Figure \ref%
{fig:descriptive}A illustrates that participation was around 65 percent in
the years before the recession at the beginning of the 1980s. Participation
drops to a sample period low of 58 percent in 1982, but subsequently
increases and almost reaches 70 percent at the end of the sample period. The
figures for the average hours show similar trends but most notably there is
significant variation in average hours over time for the sample of workers.

We employ the following variables for our empirical analysis. Three
different education levels; (1) a dummy variable indicating that the
individual left school at the age of 16 years or younger, (2) a dummy
variable for left school at the age of 17 or 18 years, and (3) a dummy
variable for left school at the age of 19 years or older. We include age and
age squared and interact these with the level of education. In addition, we
use a dummy variable indicating that the individual lives together with a
partner and 12 dummy variables indicating the region in the UK in which the
individual lives. We pool data for four consecutive years, \emph{i.e.}
1978-81, 1982-1985, 1986-1989, 1990-1993, 1994-1997 and 1998-2000, noting
that the last period is only 3 years and we assume that the model's assumptions
are satisfied for each of the pooled samples.

\pgfplotstableread{hours1.txt}{\hours} \pgfplotstableread{hours2.txt}{	%
\hoursa} \pgfplotstableread{result_part.txt}{\participation}

\begin{figure}[tbp]
	\caption{Participation rates and working hours}
	\label{fig:descriptive}
\center{
\begin{tikzpicture}[scale=0.75]
	\begin{groupplot}[
	group style={
		rows=1,
		columns=2,
		horizontal sep=50pt,
		vertical sep=50pt
	},
	xlabel={},
	ylabel={},
	xtick pos=left,
	ytick pos=left,
	width=0.6 \textwidth,
	height=0.6 \textwidth,
	ylabel = Participation rate,
	xtick = {80, 85, 90, 95},
	xtick scale label code/.code={},
	]
	\nextgroupplot[title = {A: Participation rates}]
	\addplot [black, mark = *] table{\participation};
	\nextgroupplot[title = {B: Working hours}, ylabel = Hours, ymax = 31, ymin = 8, legend style={at={(0.3,0.15)},anchor=west}]
	\addplot [black, mark = *] table{\hours};
	\addlegendentry{Full population}
	\addplot [black, mark = +] table{\hoursa}; 
	\addlegendentry{Working population}
	\end{groupplot}
	\end{tikzpicture}}
\end{figure}

\begin{table}[h]
\caption{Estimates of the returns to education without correction for sample
selection}
\label{tab:returns}\centering
{\small \ 
\begin{tabular}{l|cccc}
\hline\hline
& Mean & Q1 & Q2 & Q3 \\ \hline\hline
\multicolumn{5}{l}{Leaving school at the age of 17-18} \\ 
1978-1981 & 0.253 & 0.160 & 0.251 & 0.338 \\ 
& (0.218,0.288) & (0.122,0.198) & (0.216,0.286) & (0.223,0.452) \\ 
1986-1989 & 0.272 & 0.200 & 0.289 & 0.332 \\ 
& (0.244,0.300) & (0.169,0.231) & (0.262,0.317) & (0.265,0.399) \\ 
1998-2000 & 0.267 & 0.221 & 0.279 & 0.325 \\ 
& (0.238,0.297) & (0.183,0.258) & (0.267,0.323) & (0.268,0.382) \\ \hline
\multicolumn{5}{l}{Leaving school at the age of 19 or older} \\ 
1978-1981 & 0.640 & 0.554 & 0.715 & 0.775 \\ 
& (0.599,0.680) & (0.494,0.613) & (0.669,0.761) & (0.677,0.874) \\ 
1986-1989 & 0.567 & 0.539 & 0.686 & 0.672 \\ 
& (0.532,0.601) & (0.490,0.590) & (0.639,0.712) & (0.626,0.719) \\ 
1998-2000 & 0.593 & 0.520 & 0.687 & 0.689 \\ 
& (0.557,0.629) & (0.468, 0.572) & (0.644,0.731) & (0.644,0.732) \\ 
\hline\hline
\multicolumn{5}{l}{{\footnotesize {Bootstrapped confidence intervals shown
in parentheses}}}%
\end{tabular}
}
\end{table}

\label{ss:ret_educ}

We first examine the returns to schooling.\footnote{%
We treat the level of education as exogenous in this empirical example. Our
approach could be extended to incorporate the endogeneity of conditioning
variables but this is beyond the scope of this paper.} Table \ref%
{tab:returns} reports the impact of education on wages unadjusted for
selection. The results in the first column are the average treatment effects
of the difference between the lowest level of education and each of the two
higher levels. Similarly, columns 2 to 4 report the absolute values of the
quantile treatment effects.\footnote{%
We calculate the average treatment effects for the middle education level as
the difference between $\frac{1}{n}\sum_{i=1}w(Z_{1i},X_{i}=``middle^{\prime
\prime })^{\mathrm{T}}\widehat{\beta }$ and $\frac{1}{n}%
\sum_{i=1}w(Z_{1i},X_{i}=``low^{\prime \prime })^{\mathrm{T}}\widehat{\beta }
$, where $w$ is a polynomial and $\widehat{\beta }$ the linear-least squares
estimator. We use distribution regression for the quantile treatment effect
to estimate the distribution and calculate the quantile of that distribution.%
} The results in Table \ref{tab:returns} indicate that there is generally a
larger coefficient at higher quantiles. There is also evidence of an
increase in the return to education over time at some quantiles, although
there is no evidence of a general increase in the impact of education over
the sample period.

\captionsetup[subfigure]{labelformat=empty} 
\begin{figure}[tbp]
	\caption{Kernel density estimates of the control function.}
	\label{fig:contour}
\begin{subfigure}{\linewidth}
		\caption{\\1978-1981}
		\begin{subfigure}[b]{0.32\linewidth}
			\caption{$\leq 16$ years}
			\includegraphics[width=\linewidth]{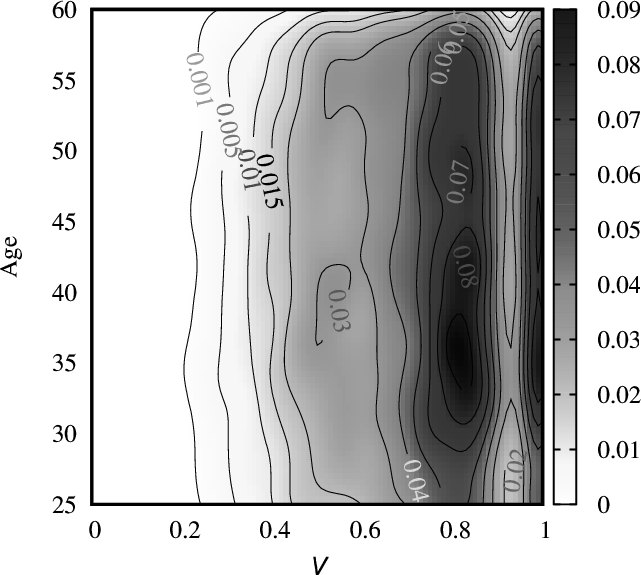}
		\end{subfigure}
		\begin{subfigure}[b]{0.32\linewidth}
			\caption{$17-18$ years}
			\includegraphics[width=\linewidth]{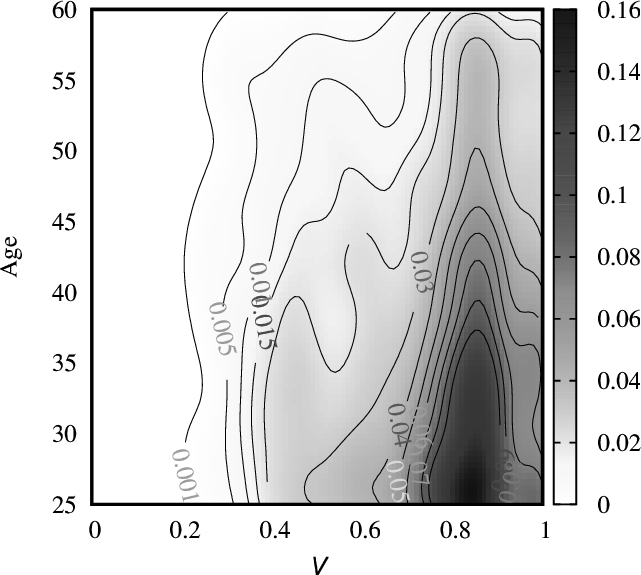}
		\end{subfigure}
		\begin{subfigure}[b]{0.32\linewidth}
			\caption{$\geq$ 19 years}
			\includegraphics[width=\linewidth]{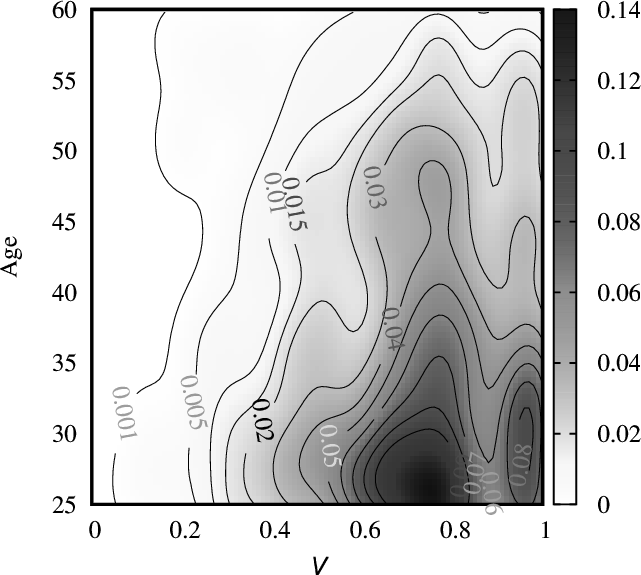}
		\end{subfigure}
	\end{subfigure}
\begin{subfigure}{\linewidth}
		\caption{\\1986-1989}
		\begin{subfigure}[b]{0.32\linewidth}
			\includegraphics[width=\linewidth]{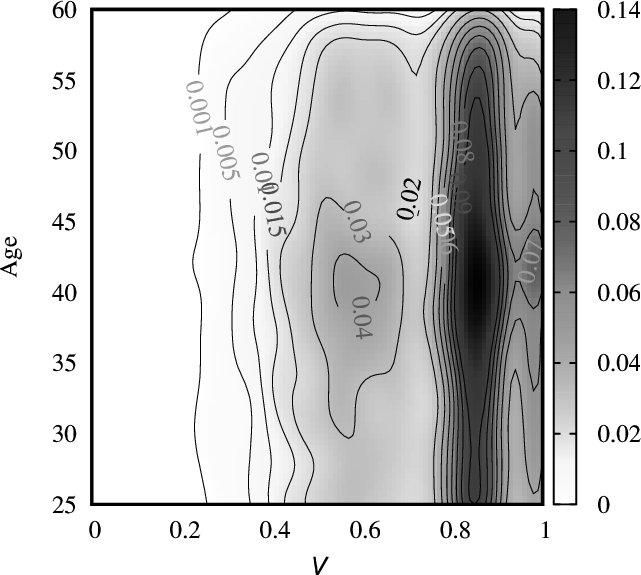}
		\end{subfigure}
		\begin{subfigure}[b]{0.32\linewidth}
			\includegraphics[width=\linewidth]{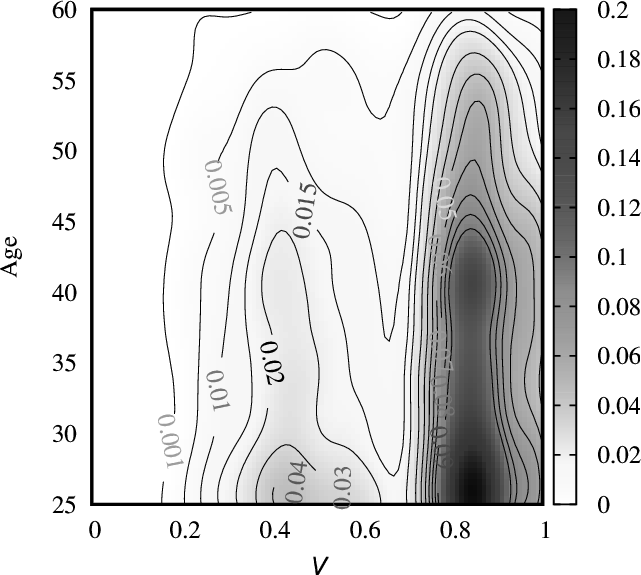}
		\end{subfigure}
		\begin{subfigure}[b]{0.32\linewidth}
			\includegraphics[width=\linewidth]{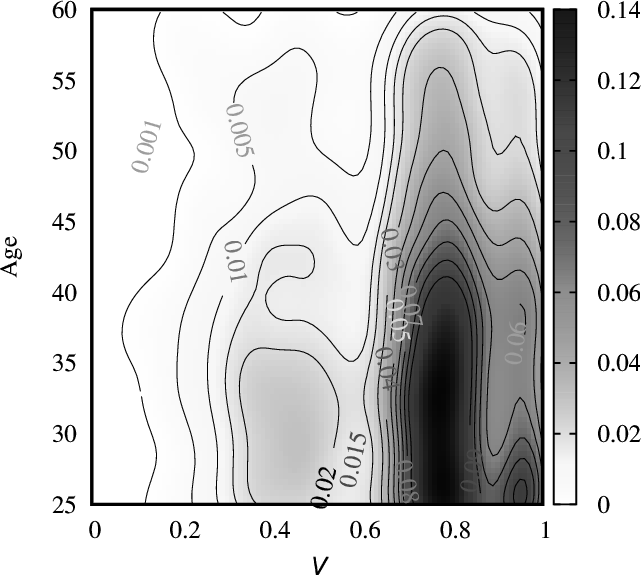}
		\end{subfigure}
	\end{subfigure}
\begin{subfigure}{\linewidth}
		\caption{\\1998-2000}
		\begin{subfigure}[b]{0.32\linewidth}
			\includegraphics[width=\linewidth]{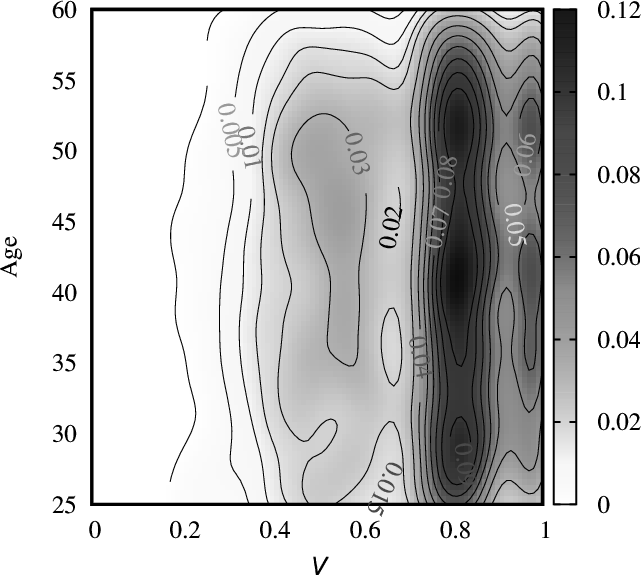}
		\end{subfigure}
		\begin{subfigure}[b]{0.32\linewidth}
			\includegraphics[width=\linewidth]{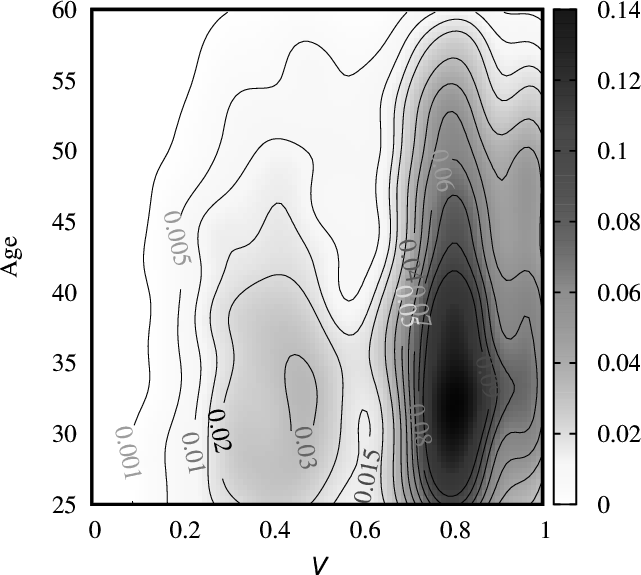}
		\end{subfigure}
		\begin{subfigure}[b]{0.32\linewidth}
			\includegraphics[width=\linewidth]{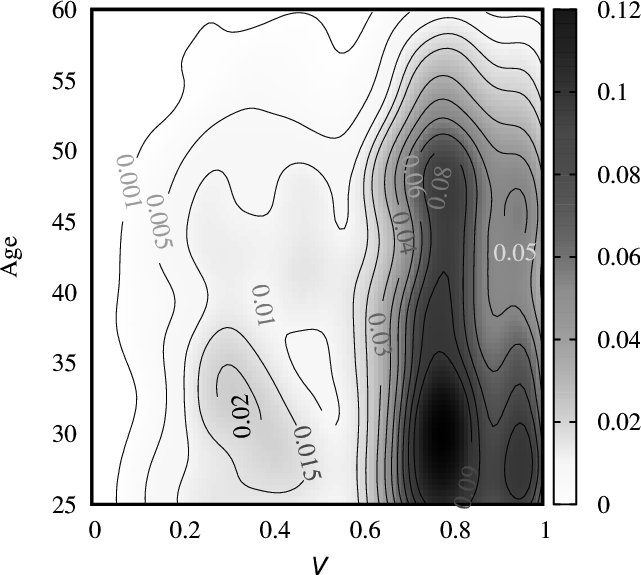}
		\end{subfigure}
	\end{subfigure}
\end{figure}

Before discussing the results produced by our approach we examine the
estimates of the control function. Figure \ref{fig:contour} presents the
joint densities of the control function and age for the different
educational categories. Age and education are the two economically most
interesting variables included in the wage equations. The nonparametric
identification of the global effects discussed in Section \ref%
{ss:integration} (Assumption \ref{assumption:cs}) requires that the supports
of the control function are identical across the education levels
considered. This appears to be satisfied and it is especially true for the
beginning of the sample period. After 1990 the support for the lowest
education level begins between 0.2 and 0.3 while that for the other
education levels begins between 0.1 and 0.2.\label{insert::kernel_discussion}

\begin{table}[tbp]
\caption{Estimates of the returns to education: Local effects adjusted for
selection.}
\label{tab:results_cf}\centering
{\small \ 
\begin{tabular}{l|cccc}
\hline\hline
& Mean & Q1 & Q2 & Q3 \\ \hline\hline
\multicolumn{5}{l}{Leaving school at the age of 17-18 years} \\ 
\multicolumn{5}{c}{$V = 0.5$} \\ \hline
1978-1981 & 0.161 & 0.158 & 0.226 & 0.300 \\ 
& (0.087, 0.221) & (0.095, 0.221) & (0.057, 0.395) & (0.045, 0.555) \\ 
1986-1989 & 0.164 & 0.126 & 0.308 & 0.426 \\ 
& (0.035, 0.169) & (0.084, 0.169) & (0.087, 0.528) & (0.055, 0.796) \\ 
1998-2000 & 0.070 & 0.184 & 0.318 & 0.411 \\ 
& (0.036, 0.242) & (0.125, 0.242) & (0.134, 0.501) & (0.144, 0.677) \\ \hline
\multicolumn{5}{c}{$V = 0.75$} \\ \hline
1978-1981 & 0.165 & 0.163 & 0.251 & 0.336 \\ 
& (0.130, 0.200) & (0.125, 0.200) & (0.127, 0.375) & (0.127, 0.545) \\ 
1986-1989 & 0.146 & 0.204 & 0.305 & 0.348 \\ 
& (0.125, 0.234) & (0.174, 0.234) & (0.183, 0.428) & (0.162, 0.534) \\ 
1998-2000 & 0.079 & 0.236 & 0.301 & 0.335 \\ 
& (0.059, 0.274) & (0.198, 0.274) & (0.210, 0.392) & (0.204, 0.465) \\ \hline
\multicolumn{5}{l}{Leaving school at the age of 19 years or older} \\ 
\multicolumn{5}{c}{$V = 0.5$} \\ \hline
1978-1981 & 0.627 & 0.609 & 0.885 & 0.954 \\ 
& (0.524, 0.765) & (0.452, 0.765) & (0.542, 1.229) & (0.589, 1.319) \\ 
1986-1989 & 0.483 & 0.458 & 0.790 & 0.877 \\ 
& (0.378, 0.603) & (0.312, 0.603) & (0.384, 1.195) & (0.436, 1.319) \\ 
1998-2000 & 0.443 & 0.395 & 0.766 & 0.890 \\ 
& (0.297, 0.495) & (0.296, 0.495) & (0.310, 1.221) & (0.354, 1.426) \\ \hline
\multicolumn{5}{c}{$V = 0.75$} \\ \hline
1978-1981 & 0.556 & 0.552 & 0.721 & 0.779 \\ 
& (0.515, 0.612) & (0.492, 0.612) & (0.509, 0.933) & (0.511, 1.047) \\ 
1986-1989 & 0.446 & 0.553 & 0.704 & 0.688 \\ 
& (0.420, 0.602) & (0.503, 0.602) & (0.522, 0.885) & (0.529, 0.846) \\ 
1998-2000 & 0.402 & 0.534 & 0.691 & 0.699 \\ 
& (0.378, 0.585) & (0.483, 0.585) & (0.505, 0.878) & (0.501, 0.896) \\ 
&  &  &  &  \\ \hline\hline
\multicolumn{5}{l}{{\footnotesize {Bootstrapped confidence intervals shown
in parentheses.}}}%
\end{tabular}
}
\end{table}

Table \ref{tab:results_cf} reports the results of the local average and
quantile treatment effects. We report the effects relative to the lowest
level of education. These local effects report the increase in terms of the
mean or quantile when we compare the wages that all individuals would
receive with the lowest level of education to those they would receive if
they had a higher level of education. These local effects correspond to the
estimates described in Section \ref{ss:local}. For example, the impact at
the mean for $V=0.5$ and \textquotedblleft Leaving school at the age of
17-18 years" is the average increase in wages of all individuals with $V=0.5$
if their education level were to increase from the lowest to the middle
education level. This impact is hypothetical since not everyone with $V=0.5$
has either a low or middle education level. With respect to the discussion
in Section \ref{ss:local}, it equals $\int \left\{ \mu (x_{1},z_{1},v)-\mu
(x_{0},z_{1},v)\right\} dF_{Z_{1}|C>0}(z_{1})$, with $x_{1}$ the middle
level of education and $x_{0}$ the lowest level of education. The quantile
treatment effects are their corresponding quantiles. We account for sample
selection by including $V$ and $V^{2}$ and the interaction of $V$ with all
the other regressors discussed above. We report results for values of $V$ at
the median and higher since it appears that our nonparametric identification
requirements are not satisfied at lower quantiles.

Table \ref{tab:results_cf} reveals that the impact of education varies by
quantile and the value of $V$ at which it is evaluated. The results for the
mean reveal some variation in the returns to education across different
values of $V$. However, the evidence is not strong statistically. This, in
addition to the similarity of the results to the unadjusted results,
suggests that there are no clear effects from selection bias.

To explicitly investigate for the presence of a selection bias, Table \ref%
{tab:average_v} provides the average derivative of wages with respect to $V$
for each of the pooled samples. This represents a test for selection bias
although a zero estimate for the average derivative does not exclude the
presence of selection since the derivative may change signs depending on the
value where it is evaluated.\footnote{%
Deriving the optimal test for selection in this setting is beyond the scope
of this paper.} The table shows that for each of the pooled samples $V$ has
a positive and statistically significant effect on wages. This indicates
that the unobservable characteristics which increase hours of work are
positively correlated with those which increase wages. Thus, there is
statistical evidence supporting the presence of selection bias and the
impact of $V$ increases over time. As the support of $V$ is generally from
0.2 to 1, a coefficient of around 0.27 indicates that the impact of $V$ on
wages is economically important.

\begin{table}[tbp]
\caption{Average derivative estimates of the impact of the control function.}
\label{tab:average_v}\centering
{\small {\ 
\begin{tabular}{lc}
\hline\hline
& Average derivative \\ \hline\hline
1978-1981 & 0.168 \\ 
& (0.136, 0.199) \\ 
1986-1989 & 0.236 \\ 
& (0.202, 0.271) \\ 
1998-2000 & 0.265 \\ 
& (0.218, 0.311) \\ \hline\hline
\end{tabular}%
} }
\end{table}

Table \ref{tab:returns1} reports the global treatment effects as presented
in Section \ref{ss:integration}.\label{insert:global_returns} For example,
the mean results for \textquotedblleft leaving school at the age of 17-18
years\textquotedblright\ is the impact on the average wage for the selected
population when their education level changes from the low to the middle
education level. From Section \ref{ss:integration}, it equals (\ref%
{eq:estimated}) for $x_{1}$ the middle level of education and $x_{0}$ the
low education level. We do not find large differences between the results
presented in Table \ref{tab:returns1} and those reported in Table \ref%
{tab:returns}. This also suggests that selection does not have an important
effect on the returns to education.

The global treatment effects are nonparametrically identified under the
strong support condition presented in Assumption \ref{assumption:cs} and as
discussed above this may not be satisfied for the later years of our
analysis. We accordingly estimate bounds for the global effects. These are
also reported in Table \ref{tab:returns1} and are tight in this particular
example. Since it is not apparent whether this reflects features of the data
or the additional information inherent in the censored selection rule, we
also 
constructed the corresponding bounds employing the propensity scores. This
did not produce informative bounds.\label{insert:bounds}

\begin{table}[tbp]
\caption{Estimates of the returns to education : Global treatment effect
adjusted for selection and bounds for global treatment effects.}
\label{tab:returns1}\centering
{\small \ 
\begin{tabular}{l|cccc}
\hline\hline
& Mean & Q1 & Q2 & Q3 \\ \hline\hline
\multicolumn{5}{l}{Leaving school at the age of 17-18 years} \\ \hline
\multicolumn{5}{l}{Estimates with $V$ as control function} \\ \hline
1978-1981 & 0.251 & 0.154 & 0.252 & 0.332 \\ 
& (0.215,0.287) & (0.113,0.199) & (0.218,0.286) & (0.293,0.371) \\ 
&  & [0.131,0.182] & [0.231,0.279] & [0.301,0.370] \\ 
1986-1989 & 0.286 & 0.217 & 0.304 & 0.346 \\ 
& (0.257,0.314) & (0.183,0.249) & (0.278,0.329) & (0.303,0.389) \\ 
&  & [0.194,0.245] & [0.267,0.324] & [0.277,0.372] \\ 
1998-2000 & 0.285 & 0.247 & 0.315 & 0.339 \\ 
& (0.256,0.314) & (0.195,0.297) & (0.271,0.359) & (0.297,0.381) \\ 
&  & [0.211,0.293] & [0.263,0.341] & [0.245,0.363] \\ \hline
\multicolumn{5}{l}{Leaving school at the age of 19 years or older} \\ \hline
\multicolumn{5}{l}{Estimates with $V$ as control function} \\ \hline
1978-1981 & 0.632 & 0.534 & 0.695 & 0.765 \\ 
& (0.591,0.672) & (0.477,0.592) & (0.645,0.745) & (0.714,0.816) \\ 
&  & [0.521,0.550] & [0.680,0.707] & [0.738,0.780] \\ 
1986-1989 & 0.574 & 0.554 & 0.694 & 0.674 \\ 
& (0.541,0.607) & (0.506,0.602) & (0.656,0.731) & (0.646,0.701) \\ 
&  & [0.533,0.582] & [0.656,0.710] & [0.612,0.686] \\ 
1998-2000 & 0.604 & 0.541 & 0.694 & 0.689 \\ 
& (0.570,0.639) & (0.482,0.599) & (0.644,0.744) & (0.644,0.734) \\ 
&  & [0.506,0.584] & [0.642,0.718] & [0.593,0.704] \\ \hline\hline
\multicolumn{5}{p{11cm}}{{\footnotesize {Bootstrapped confidence intervals
are in parentheses. Bounds are in square brackets.}}}%
\end{tabular}
}
\end{table}

\pgfplotstableread{global_5yrs_new/src/results_global1_1_5.txt}{\resultsa} %
\pgfplotstableread{global_5yrs_new/src/results_global1_2_5.txt}{\resultsb} %
\pgfplotstableread{global_5yrs_new/src/results_global1_3_5.txt}{\resultsc} %
\pgfplotstableread{global_5yrs_new/src/results_global2_1_5.txt}{\resultsaa} %
\pgfplotstableread{global_5yrs_new/src/results_global2_2_5.txt}{\resultsba} %
\pgfplotstableread{global_5yrs_new/src/results_global2_3_5.txt}{\resultsca} %
\pgfplotstableread{global_5yrs_new/src/results_global3_1_5.txt}{\resultsab} %
\pgfplotstableread{global_5yrs_new/src/results_global3_2_5.txt}{\resultsbb} %
\pgfplotstableread{global_5yrs_new/src/results_global3_3_5.txt}{\resultscb}

\begin{figure}[tbp]
\renewcommand{\thefigure}{\arabic{figure}A} 
\caption{Global estimates of the average impact of education for the low and
	middle educated in the selected population when assigned an alternative
	education level.}
\label{fig:att}
\begin{tikzpicture}
	\begin{groupplot}
	[
	group style={%
		group size = 3 by 1,
		group name=plots
		, horizontal sep=0.5cm,
		y descriptions at=edge left,
		ylabels at=edge left,
	},
	set layers,cell picture=true,
	width=0.37 \textwidth,
	height=0.37 \textwidth,
	legend columns=-1,
	xlabel = Year,
	ylabel = Difference,
	ymin = 0,
	ymax = 1.,
	xtick = {80, 85, 90, 95 }, 
	cycle list name=black white
	]
	\nextgroupplot[legend to name=grouplegend2, title = {Low vs. middle}]
	\addplot[name path = a, black] table{\resultsa};		
	\addplot[very thick, black] table{\resultsb};
	\addplot[name path = b, black] table{\resultsc};
	\addplot[gray!30] fill between[ 
	of = a and b
	]; 
	\nextgroupplot[title = {High vs. low}]
	\addplot[name path = a, black] table{\resultsaa};		
	\addplot[very thick, black] table{\resultsba};
	\addplot[name path = b, black] table{\resultsca};
	\addplot[gray!30] fill between[ 
	of = a and b
	]; 
	\nextgroupplot[title = {High vs. middle} ]
	\addplot[name path = a, black] table{\resultsab};		
	\addplot[very thick, black] table{\resultsbb};
	\addplot[name path = b, black] table{\resultscb};
	\addplot[gray!30] fill between[ 
	of = a and b
	]; 
	\end{groupplot}
	\end{tikzpicture}
\end{figure}

We also explore the influence of education by deriving the average and
quantile impacts of obtaining a higher level of education for specific
groups. These are the treatment effects for the treated discussed in Section %
\ref{ss:global_treated} and are shown in Figures \ref{fig:att} and \ref%
{fig:median_treatment}. The labels \textquotedblleft Low\textquotedblright ,
\textquotedblleft Middle\textquotedblright\ and \textquotedblleft
High\textquotedblright\ capture the three education groups. These effects
are nonparametrically identified under the weak common support condition in
Assumption \ref{assumption:wcs}. This requirement seems satisfied as the
support of the control function for the ``Middle'' and ``High'' education
levels are at least as large as that of the education level ``Low'', while
the support of the control function for the education level ``High'' is at
least as large as that of the education level ``Medium'' (see Figure \ref%
{fig:contour}). The estimates are based on pooling the data as described
above. The figure \textquotedblleft Low versus middle\textquotedblright\ in
Figure \ref{fig:att} displays the average increase in wages when individuals
of the low education group have an education level equal to the middle
education level. These are the estimates described in Section \ref%
{ss:global_treated} and correspond to the marginal increase in the education
attainment. In particular, we estimate \eqref{eq:estimated1}. These contrast
to those in Table \ref{tab:returns1} which report the impact on the mean or
quantiles if everyone in the selected sample went from a low to a higher
education level, irrespective of the observed education levels of these
individuals. The figure displayed for $\tau =0.25$ in Figure \ref%
{fig:median_treatment} presents this marginal increase at the first quartile
of the distribution. The magnitude of the average impact of education for
the various educational comparisons is consistent with the estimates
discussed above and the plot over time appears to reveal some cyclical
behavior.

\pgfplotstableread{global1_5yrs_new/src/results_global1_1_25.txt}{\resultsa} %
\pgfplotstableread{global1_5yrs_new/src/results_global1_2_25.txt}{\resultsb} %
\pgfplotstableread{global1_5yrs_new/src/results_global1_3_25.txt}{\resultsc} %
\pgfplotstableread{global1_5yrs_new/src/results_global2_1_25.txt}{\resultsaa}
\pgfplotstableread{global1_5yrs_new/src/results_global2_2_25.txt}{\resultsba}
\pgfplotstableread{global1_5yrs_new/src/results_global2_3_25.txt}{\resultsca}
\pgfplotstableread{global1_5yrs_new/src/results_global3_1_25.txt}{\resultsab}
\pgfplotstableread{global1_5yrs_new/src/results_global3_2_25.txt}{\resultsbb}
\pgfplotstableread{global1_5yrs_new/src/results_global3_3_25.txt}{\resultscb}

\pgfplotstableread{global1_5yrs_new/src/results_global1_1_5.txt}{\resultsac} %
\pgfplotstableread{global1_5yrs_new/src/results_global1_2_5.txt}{\resultsbc} %
\pgfplotstableread{global1_5yrs_new/src/results_global1_3_5.txt}{\resultscc} %
\pgfplotstableread{global1_5yrs_new/src/results_global2_1_5.txt}{\resultsad} %
\pgfplotstableread{global1_5yrs_new/src/results_global2_2_5.txt}{\resultsbd} %
\pgfplotstableread{global1_5yrs_new/src/results_global2_3_5.txt}{\resultscd} %
\pgfplotstableread{global1_5yrs_new/src/results_global3_1_5.txt}{\resultsae} %
\pgfplotstableread{global1_5yrs_new/src/results_global3_2_5.txt}{\resultsbe} %
\pgfplotstableread{global1_5yrs_new/src/results_global3_3_5.txt}{\resultsce}

\pgfplotstableread{global1_5yrs_new/src/results_global1_1_75.txt}{\resultsaf}
\pgfplotstableread{global1_5yrs_new/src/results_global1_2_75.txt}{\resultsbf}
\pgfplotstableread{global1_5yrs_new/src/results_global1_3_75.txt}{\resultscf}
\pgfplotstableread{global1_5yrs_new/src/results_global2_1_75.txt}{\resultsag}
\pgfplotstableread{global1_5yrs_new/src/results_global2_2_75.txt}{\resultsbg}
\pgfplotstableread{global1_5yrs_new/src/results_global2_3_75.txt}{\resultscg}
\pgfplotstableread{global1_5yrs_new/src/results_global3_1_75.txt}{\resultsah}
\pgfplotstableread{global1_5yrs_new/src/results_global3_2_75.txt}{\resultsbh}
\pgfplotstableread{global1_5yrs_new/src/results_global3_3_75.txt}{\resultsch}

\begin{figure}[tbp]
\addtocounter{figure}{-1} \renewcommand{\thefigure}{\arabic{figure}B} 
\caption{Global estimates of the quantile impact of education for the low
	and middle educated in the selected population when assigned an alternative
	education level.}
\label{fig:median_treatment}
\begin{tikzpicture}
	\begin{groupplot}
	[
	group style={%
		group size = 3 by 3
		, horizontal sep=0.5cm, 
		vertical sep=1cm,
		group name=plots,
		xlabels at=edge bottom,
		y descriptions at=edge left,
		ylabels at=edge left,
		x descriptions at=edge bottom
	},
	set layers,cell picture=true,
	width=0.37\textwidth,
	height=0.37 \textwidth,
	legend columns=-1,
	xlabel = Year,
	ylabel = Difference,
	ymin = 0,
	ymax = 1.,
	xtick = {80, 85, 90, 95 }, 
	cycle list name=black white
	]
	\nextgroupplot[legend to name=grouplegend3]
	\addplot[name path = a, black] table{\resultsa};		
	\addplot[very thick, black] table{\resultsb};
	\addplot[name path = b, black] table{\resultsc};
	\addplot[gray!30] fill between[ 
	of = a and b
	]; 
	\nextgroupplot[]
	\addplot[name path = a, black] table{\resultsaa};		
	\addplot[very thick, black] table{\resultsba};
	\addplot[name path = b, black] table{\resultsca};
	\addplot[gray!30] fill between[ 
	of = a and b
	]; 
	\nextgroupplot[]
	\addplot[name path = a, black] table{\resultsab};		
	\addplot[very thick, black] table{\resultsbb};
	\addplot[name path = b, black] table{\resultscb};
	\addplot[gray!30] fill between[ 
	of = a and b
	]; 
	\nextgroupplot[legend to name=grouplegend]
	\addplot[name path = a, black] table{\resultsac};		
	\addplot[very thick, black] table{\resultsbc};
	\addplot[name path = b, black] table{\resultscc};
	\addplot[gray!30] fill between[ 
	of = a and b
	]; 
	\nextgroupplot[]
	\addplot[name path = a, black] table{\resultsad};		
	\addplot[very thick, black] table{\resultsbd};
	\addplot[name path = b, black] table{\resultscd};
	\addplot[gray!30] fill between[ 
	of = a and b
	]; 
	\nextgroupplot[]
	\addplot[name path = a, black] table{\resultsae};		
	\addplot[very thick, black] table{\resultsbe};
	\addplot[name path = b, black] table{\resultsce};
	\addplot[gray!30] fill between[ 
	of = a and b
	]; 
	\nextgroupplot[]
	\addplot[name path = a, black] table{\resultsaf};		
	\addplot[very thick, black] table{\resultsbf};
	\addplot[name path = b, black] table{\resultscf};
	\addplot[gray!30] fill between[ 
	of = a and b
	]; 
	\nextgroupplot[]
	\addplot[name path = a, black] table{\resultsag};		
	\addplot[very thick, black] table{\resultsbg};
	\addplot[name path = b, black] table{\resultscg};
	\addplot[gray!30] fill between[ 
	of = a and b
	]; 
	\nextgroupplot[]
	\addplot[name path = a, black] table{\resultsah};		
	\addplot[very thick, black] table{\resultsbh};
	\addplot[name path = b, black] table{\resultsch};
	\addplot[gray!30] fill between[ 
	of = a and b
	]; 
	\end{groupplot}
	\node at (plots c1r1.north) [yshift=1cm] {Low vs. Middle};  
	\node at (plots c2r1.north) [yshift=1cm] {Low vs. High};  
	\node at (plots c3r1.north) [yshift=1cm] {Middle vs. High};  
	\node at (plots c2r1.north) [yshift=0.3cm] {$\tau = 0.25$}; 
	\node at (plots c2r2.north) [yshift=0.3cm] {$\tau = 0.50$}; '
	\node at (plots c2r3.north) [yshift=0.3cm] {$\tau = 0.75$};  
	\end{tikzpicture}
\end{figure}

The different estimates of the impact of education may partially capture
that the returns to education vary by birth cohort. For example, the
increasing numbers of college educated females may reflect that selection
into such education has changed (i.e. has become less or more demanding) and
may result in lower or higher returns to education over time. We investigate
this by estimating birth cohort specific age profiles. We find that older
cohorts have slightly higher wages but that the differences are small.\label%
{insert:cohorts}

We also explore how the return to experience has varied by education group
over the sample period by estimating the average derivative with respect to
age. Figure \ref{fig:average} presents the derivative for different
education levels. The figures show that there is a drastic increase to the
return to experience during the 1990s. They also reveal that there is a
drastic difference in the rate of wage growth across education groups.
Figure \ref{fig:local_average_response} reports these derivatives evaluated
at ages 25, 40, and 55 years and these represent the local average
responses. There is a strong positive relationship between wage growth and
age at 25 years and the effect is particularly strong for the highest
educated. Moreover, the effect increases notably over the sample period with
large increases in the 1990s. The effect is notably lower although still
positive at the age of 40 years. The differences by education groups are
less dramatic. At the age of 55 years wages do not appear to be generally
increasing with age. In fact, there appears to be evidence that the real
wage is decreasing for the highest education group.

\pgfplotstableread{average_derivative/src/results_global1_1_5.txt}{\resultsa}
\pgfplotstableread{average_derivative/src/results_global1_2_5.txt}{\resultsb}
\pgfplotstableread{average_derivative/src/results_global1_3_5.txt}{\resultsc}
\pgfplotstableread{average_derivative/src/results_global2_1_5.txt}{		%
\resultsaa} %
\pgfplotstableread{average_derivative/src/results_global2_2_5.txt}{		%
\resultsba} %
\pgfplotstableread{average_derivative/src/results_global2_3_5.txt}{		%
\resultsca} %
\pgfplotstableread{average_derivative/src/results_global3_1_5.txt}{		%
\resultsab} %
\pgfplotstableread{average_derivative/src/results_global3_2_5.txt}{		%
\resultsbb} %
\pgfplotstableread{average_derivative/src/results_global3_3_5.txt}{		%
\resultscb} %
\pgfplotstableread{average_derivative/src/results_global4_1_5.txt}{		%
\resultsac} %
\pgfplotstableread{average_derivative/src/results_global4_2_5.txt}{		%
\resultsbc} %
\pgfplotstableread{average_derivative/src/results_global4_3_5.txt}{		%
\resultscc}

\begin{figure}[h]
	\caption{Average derivative of the impact of age on log wages among the
		selected population.}
	\label{fig:average}
\begin{tikzpicture}
	\begin{groupplot}
	[
	group style={%
		group size = 3 by 1,
		group name=plots,
		xlabels at=edge bottom,
		horizontal sep=0.5cm, 
		y descriptions at=edge left,
		ylabels at=edge left,
		x descriptions at=edge bottom
	},
	set layers,cell picture=true,
	scaled ticks=false, tick label style={/pgf/number format/fixed},
	width=0.37 \textwidth,
	height=0.37 \textwidth,
	legend columns=-1,
	xlabel = Year,
	ylabel = Average derivative,
	ymin = -0.01,
	ymax = 0.03,     
	xtick = {80, 85, 90, 95 },  
	cycle list name=black white
	]
	\nextgroupplot[title = {A. Low}]
	\addplot[name path = a, black] table{\resultsaa};		
	\addplot[very thick, black] table{\resultsba};
	\addplot[name path = b, black] table{\resultsca};
	\addplot[gray!30] fill between[ 
	of = a and b
	]; 
	\nextgroupplot[title = {B. Middle}]
	\addplot[name path = a, black] table{\resultsab};		
	\addplot[very thick, black] table{\resultsbb};
	\addplot[name path = b, black] table{\resultscb};
	\addplot[gray!30] fill between[ 
	of = a and b
	]; 
	\nextgroupplot[title = {C. High}]
	\addplot[name path = a, black] table{\resultsac};		
	\addplot[very thick, black] table{\resultsbc};
	\addplot[name path = b, black] table{\resultscc};
	\addplot[gray!30] fill between[ 
	of = a and b
	]; 
	\end{groupplot}
	\end{tikzpicture}
\end{figure}

\pgfplotstableread{local_average_response/src/results_global1_1_5_25.txt}{		%
\resultsa} %
\pgfplotstableread{local_average_response/src/results_global1_2_5_25.txt}{		%
\resultsb} %
\pgfplotstableread{local_average_response/src/results_global1_3_5_25.txt}{		%
\resultsc} %
\pgfplotstableread{local_average_response/src/results_global2_1_5_25.txt}{		%
\resultsaa} %
\pgfplotstableread{local_average_response/src/results_global2_2_5_25.txt}{		%
\resultsba} %
\pgfplotstableread{local_average_response/src/results_global2_3_5_25.txt}{		%
\resultsca} %
\pgfplotstableread{local_average_response/src/results_global3_1_5_25.txt}{		%
\resultsab} %
\pgfplotstableread{local_average_response/src/results_global3_2_5_25.txt}{		%
\resultsbb} %
\pgfplotstableread{local_average_response/src/results_global3_3_5_25.txt}{		%
\resultscb} %
\pgfplotstableread{local_average_response/src/results_global4_1_5_25.txt}{		%
\resultsac} %
\pgfplotstableread{local_average_response/src/results_global4_2_5_25.txt}{		%
\resultsbc} %
\pgfplotstableread{local_average_response/src/results_global4_3_5_25.txt}{		%
\resultscc}

\pgfplotstableread{local_average_response/src/results_global1_1_5_40.txt}{		%
\resultsamiddle} %
\pgfplotstableread{local_average_response/src/results_global1_2_5_40.txt}{		%
\resultsbmiddle} %
\pgfplotstableread{local_average_response/src/results_global1_3_5_40.txt}{		%
\resultscmiddle} %
\pgfplotstableread{local_average_response/src/results_global2_1_5_40.txt}{		%
\resultsaamiddle} %
\pgfplotstableread{local_average_response/src/results_global2_2_5_40.txt}{		%
\resultsbamiddle} %
\pgfplotstableread{local_average_response/src/results_global2_3_5_40.txt}{		%
\resultscamiddle} %
\pgfplotstableread{local_average_response/src/results_global3_1_5_40.txt}{		%
\resultsabmiddle} %
\pgfplotstableread{local_average_response/src/results_global3_2_5_40.txt}{		%
\resultsbbmiddle} %
\pgfplotstableread{local_average_response/src/results_global3_3_5_40.txt}{		%
\resultscbmiddle} %
\pgfplotstableread{local_average_response/src/results_global4_1_5_40.txt}{		%
\resultsacmiddle} %
\pgfplotstableread{local_average_response/src/results_global4_2_5_40.txt}{		%
\resultsbcmiddle} %
\pgfplotstableread{local_average_response/src/results_global4_3_5_40.txt}{		%
\resultsccmiddle}

\pgfplotstableread{local_average_response/src/results_global1_1_5_55.txt}{		%
\resultsahighest} %
\pgfplotstableread{local_average_response/src/results_global1_2_5_55.txt}{		%
\resultsbhighest} %
\pgfplotstableread{local_average_response/src/results_global1_3_5_55.txt}{		%
\resultschighest} %
\pgfplotstableread{local_average_response/src/results_global2_1_5_55.txt}{		%
\resultsaahighest} %
\pgfplotstableread{local_average_response/src/results_global2_2_5_55.txt}{		%
\resultsbahighest} %
\pgfplotstableread{local_average_response/src/results_global2_3_5_55.txt}{		%
\resultscahighest} %
\pgfplotstableread{local_average_response/src/results_global3_1_5_55.txt}{		%
\resultsabhighest} %
\pgfplotstableread{local_average_response/src/results_global3_2_5_55.txt}{		%
\resultsbbhighest} %
\pgfplotstableread{local_average_response/src/results_global3_3_5_55.txt}{		%
\resultscbhighest} %
\pgfplotstableread{local_average_response/src/results_global4_1_5_55.txt}{		%
\resultsachighest} %
\pgfplotstableread{local_average_response/src/results_global4_2_5_55.txt}{		%
\resultsbchighest} %
\pgfplotstableread{local_average_response/src/results_global4_3_5_55.txt}{		%
\resultscchighest}

\begin{figure}[h!]
	\caption{Local average response at different ages among the selected
		population.}
	\label{fig:local_average_response}
\begin{tikzpicture}
	\begin{groupplot}
	[
	group style={%
		group size = 3 by 3,
		group name=plots,
		vertical sep=2cm
		, horizontal sep=0.5cm, 
		xlabels at=edge bottom,
		y descriptions at=edge left,
		ylabels at=edge left,
		x descriptions at=edge bottom
	},
	set layers,cell picture=true,
	scaled ticks=false, tick label style={/pgf/number format/fixed},
	width=0.37 \textwidth,
	height=0.37 \textwidth,
	legend columns=-1,
	xlabel = Year,
	ylabel = Local average response,
	ymin = -0.05,
	ymax = 0.06,     
	xtick = {80, 85, 90, 95}, 
	cycle list name=black white
	]
	\nextgroupplot[title = {Low}]
	\addplot[name path = a, black] table{\resultsaa};		
	\addplot[very thick, black] table{\resultsba};
	\addplot[name path = b, black] table{\resultsca};
	\addplot[gray!30] fill between[ 
	of = a and b
	]; 
	\nextgroupplot[title = {Middle}]
	\addplot[name path = a, black] table{\resultsab};		
	\addplot[very thick, black] table{\resultsbb};
	\addplot[name path = b, black] table{\resultscb};
	\addplot[gray!30] fill between[ 
	of = a and b
	]; 
	\nextgroupplot[title = {High}]
	\addplot[name path = a, black] table{\resultsac};		
	\addplot[very thick, black] table{\resultsbc};
	\addplot[name path = b, black] table{\resultscc};
	\addplot[gray!30] fill between[ 
	of = a and b
	]; 
\nextgroupplot
\addplot[name path = a, black] table{\resultsaamiddle};		
\addplot[very thick, black] table{\resultsbamiddle};
\addplot[name path = b, black] table{\resultscamiddle};
\addplot[gray!30] fill between[ 
of = a and b
]; 
\nextgroupplot
\addplot[name path = a, black] table{\resultsabmiddle};		
\addplot[very thick, black] table{\resultsbbmiddle};
\addplot[name path = b, black] table{\resultscbmiddle};
\addplot[gray!30] fill between[ 
of = a and b
]; 
\nextgroupplot
\addplot[name path = a, black] table{\resultsacmiddle};		
\addplot[very thick, black] table{\resultsbcmiddle};
\addplot[name path = b, black] table{\resultsccmiddle};
\addplot[gray!30] fill between[ 
of = a and b
]; 
\nextgroupplot
\addplot[name path = a, black] table{\resultsaahighest};		
\addplot[very thick, black] table{\resultsbahighest};
\addplot[name path = b, black] table{\resultscahighest};
\addplot[gray!30] fill between[ 
of = a and b
]; 
\nextgroupplot
\addplot[name path = a, black] table{\resultsabhighest};		
\addplot[very thick, black] table{\resultsbbhighest};
\addplot[name path = b, black] table{\resultscbhighest};
\addplot[gray!30] fill between[ 
of = a and b
]; 
\nextgroupplot
\addplot[name path = a, black] table{\resultsachighest};		
\addplot[very thick, black] table{\resultsbchighest};
\addplot[name path = b, black] table{\resultscchighest};
\addplot[gray!30] fill between[ 
of = a and b
]; 
\end{groupplot}
\node at (plots c2r1.north) [yshift=1.25cm] {25 years}; 
\node at (plots c2r2.north) [yshift=1.25cm] {40 years}; 
\node at (plots c2r3.north) [yshift=1.25cm] {55 years};  
\end{tikzpicture}
\end{figure}

This empirical investigation uncovers a number of interesting features of
the data and highlights several benefits of our approach. First, despite the
large change in the participation rates of females over this period there
are no corresponding changes in the return to education. However, there is
evidence of an increase in the return to experience in the 1990s. Our
approach uncovers the presence of a selection bias and although this does
not appear to influence the estimates of the return to education, there is
an increase in the return to unobservables which influence hours over the
sample period. Finally, through this empirical example we have illustrated
the substantial reduction in bounds which results from the use of the
control function associated with a censored selection rule in contrast to
those from the use of the propensity score associated with a binary
selection rule.

\section{Conclusion}

This paper examines a nonseparable sample selection model with a selection
equation which is based on a censored outcome. We account for selection by
conditioning on an appropriately constructed control function. We show that
we are able to identify several economically interesting objects. We
categorize these as local effects which represent estimands conditional on a
specific outcome of the control function and global effects which represent
estimands evaluated over a range of values of the control function. For both
effects we provide identification results and estimation methods and the
related asymptotic theory. We illustrate the utility of our approach in an
empirical application focusing on the determinants of wages for a sample of
females in the UK over a period of increasing labor force participation at
both the intensive and extensive margins.

We provide $\sqrt{n}$-consistent estimators of the effects of interest under
the correct semiparametric specification of the control function. This
specification could also be used as the basis for sieve methods that are
more robust to functional form assumptions (see, for example, Chen 2007).
Analyzing the asymptotic properties of the resulting estimators whose
dimensions grow with the sample size would require a different proof
strategy. We rely on weak dependence of the estimator of the control
function to apply the delta method, which is not available for sieve
estimators. It might be possible to derive theory for sieve estimators by
using different techniques such as strong approximations. This is a useful
extension that we leave to future work.



\appendix
\linespread{0.98}

\section{{\protect\small Proofs of Section \protect\ref{sec:identification}}}

\subsection{{\protect\small Lemma \protect\ref{lem:cv}}}

{\small \label{app:cv} }

\begin{proof}
{\small The proof is similar to the proof of Theorem 1 in Newey (2007). For
any bounded function $a(\varepsilon)$ and $C>0$ (and hence $h(Z,\eta) > 0$),
by Assumption \ref{assumption:cv}, 
\begin{equation*}
\mathbb{E}\left[ a(\varepsilon ) \mid Z=z,\eta=q, C>0 \right] = \mathbb{E}%
\left[ a(\varepsilon ) \mid \eta=q, C>0 \right]
\end{equation*}
Since this holds for any function $a(\varepsilon )$ and any $h(Z,\eta) > 0 $%
, $Z$ and $\varepsilon $ are independent conditional on $\eta $ and $C>0$.
The result follows because $\eta $ is a one-to-one function of $F_{C\mid
Z}(C\mid Z)$ when $C>0$ since $\eta =h^{-1}(Z,C)$ if $C>0$ by Assumption \ref%
{assumption:cv}, and for $c>0$ 
\begin{eqnarray*}
F_{C\mid Z}(c \mid Z=z)&=& P(\max (h(Z,\eta ),0)\leq c\mid Z=z) \\
&=& P(h(Z,\eta ) \leq c\mid Z = z)= P(\eta \leq h^{-1}(Z,c)\mid Z)=
h^{-1}(Z,c),
\end{eqnarray*}%
where we use the normalization $\eta \sim U(0,1)$. 
}
\end{proof}

\subsection{{\protect\small Theorem \protect\ref{lem:ident_d_star}}}

{\small \label{app:ident_d_star} }

\begin{proof}
{\small Define the generic local object 
\begin{equation*}
\theta(x,v) = \mathbb{E}_{\varepsilon}[\Gamma(x,z_1,\varepsilon)\mid V=v]
\end{equation*}
for some function $\Gamma(x,z_1,e): \mathcal{X} \times \mathcal{Z}_1 \times 
\mathcal{E} \rightarrow \mathbb{R}^k; k \in \mathbb{N}^+$, where $\mathcal{E}
$ is the support of $\varepsilon$. Using Assumption \ref{assumption:cv},
this equals 
\begin{equation*}
\theta(x,z_1,v) = \mathbb{E}_{\varepsilon}[\Gamma(x,z_1,\varepsilon)\mid Z =
z, V=v].
\end{equation*}
Since conditional on $Z=z$ and $V=v$, we have that $C = \max\{h(z,v), 0\}$
and since $(x,z_1,v) \in \mathcal{XZ}_1\mathcal{V}$, there is a $z \in Z$
such that $C = h(z,v) > 0$ and hence 
\begin{equation*}
\begin{split}
\theta(x,z_1,v) & = \mathbb{E}_{\varepsilon}[\Gamma(x,z_1,\varepsilon)\mid Z
= z, V=v, h(z,v)>0] \\
& = \mathbb{E}_{\varepsilon}[\Gamma(x,z_1,\varepsilon)\mid Z = z, V=v, C>0]
\\
& = \mathbb{E}_{\varepsilon}[\Gamma(X,Z_1,\varepsilon)\mid Z = z, V=v, C>0],
\end{split}%
\end{equation*}
where the third line is due to $(X,Z_1) \subseteq Z$. This third line is
identical to the right-hand side of (\ref{eq:asf_identified}) when $%
\Gamma(x,z_1,e) = g(x,z_1,e)$. Along the same lines, this also proves
Corollary \ref{theorem:local_average_derivative}, with $\Gamma(x,z_1,e) =
\partial_x g(x,z_1,e)$. Note that this also proves the identification of the
LDSF, since 
\begin{equation*}
G(y,x,z_1,v) = \mathbb{E}_{\varepsilon}[\mathbf{1}\{ g(x,z_1,\varepsilon)
\leq y\}\mid V=v]
\end{equation*}
and hence the proof is completed by using $\Gamma(x,z_1,e) = \mathbf{1}\{
g(x,z_1,e) \leq y\}$. }
\end{proof}

{\small 
}

{\small 
}

\subsection{{\protect\small {Proof of tighter bounds using censored
selection model rather than dichotomous selection}}}

\label{app:tighter_bounds}

{\small {We drop $Z_1$ here to simplify the notation. To show that using the
control function $V,$ rather than the propensity score $P = 1 - F_{C}( 0
\mid Z),$ produces narrower bounds, we demonstrate the sufficient but not
necessary condition that the event $\{V\in \bar{\mathcal{V}}(x)\}$ implies
the event $\{P\in \bar{\mathcal{P}}(x)\}$, conditional on $C>0$, where $\bar{%
\mathcal{P}}(x) = \mathcal{P} \setminus \mathcal{P}(x),$ $\mathcal{P}$ is
the support of $P$ conditional on $C>0$, and $\mathcal{P}(x)$ is the support
of $P$ conditional on $X=x$ and $C>0$. We assume that all supports are
connected intervals, $\mathcal{P}=[\underline{p},\overline{p}]$, $\mathcal{P}%
(x)=[\underline{p}(x),\overline{p}(x)]$, $\mathcal{V}=[\underline{v},%
\overline{v}] $, and $\mathcal{V}(x)=[\underline{v}(x),\overline{v}(x)]$.
Let $Z=(X,\tilde Z)$ and $z=(x,\tilde z)$. We first show that $\underline{v}%
(x)>\underline{v}$ implies that $\overline{p}(x)<\overline{p}$. Note that $%
\underline{v}(x)>\underline{v}$ if there exists $x_0\in \mathcal{X}$ such
that $\inf \{F_{C}(c\mid x,\tilde z):(c,\tilde z)\in \mathrm{supp}(C,\tilde
Z\mid X=x,C>0)\}>\inf \{F_{C}(c\mid x_0,\tilde z):(c,\tilde z)\in \mathrm{%
supp}(C,\tilde Z \mid X=x_0,C>0)\}$. Since $c\mapsto \{F_{C}(c\mid x,\tilde
z)$ is increasing for any $x$, this implies that $\sup \{1-F_{C}(0\mid
x,\tilde z):\tilde z \in \mathrm{supp}(\tilde Z \mid X=x,C>0)\}<\sup
\{1-F_{C}(0\mid x_0,\tilde z):\tilde z \in \mathrm{supp}(\tilde Z \mid
X=x_0,C>0)\}$ as $c\mapsto F_{C}(c\mid X,\tilde Z)$ is continuous from the
right at $c=0$ almost surely, and therefore $\overline{p}(x)>\overline{p}$.
To complete the proof, we show that $\overline{v}(x)=\overline{v}=1$, so
that $\{\overline{v}(x)<\overline{v}\}=\emptyset $ necessarily implies $\{%
\underline{p}(x)>\underline{p}\}$. The result follows from $\overline{v}%
(x)=\sup \{F_{C}(c\mid x,\tilde z):(c,\tilde z)\in \mathrm{supp}(C,\tilde Z
\mid X=x,C>0)\}=1$ because $c\mapsto F_{C}(c\mid x,\tilde z)$ is a
distribution function.} }

\section{{\protect\small Proofs of Section \protect\ref{sec:estimation}}}

{\small \label{app:estimation} }

\subsection{\protect\small Notation}

{\small \label{app:notation} In what follows, $\vartheta$ denotes a generic
value of the control function. It is also convenient to introduce some
additional notation, which will be extensively used in the proofs. Let $%
V_i(\vartheta) := \vartheta(Z_{i})$, $W_i(\vartheta) := w( X_{i},Z_{1i},
V_i(\vartheta))$, and $\dot{W}_i(\vartheta) := \partial_v w(X_{i},Z_{1i},
v)|_{v = V_i(\vartheta)}$. When the previous functions are evaluated at the
true values we use $V_i = V_i(\vartheta_0),$ $W_i = W_i(\vartheta_0)$, and $%
\dot{W}_i = \dot{W}_i(\vartheta_0)$. 
Recall that $A := (Y*1(C>0),C,Z, V)$, $T(c) = 1(c \in \overline{\mathcal{C}}%
), $ and $T = T(C)$. For a function $f: \mathcal{A} \mapsto \mathbb{R}$, we
use $\| f \|_{T,\infty} = \sup_{a \in \mathcal{A}}|T(c) f(a)|$; for a $K$%
-vector of functions $f: \mathcal{A} \mapsto \mathbb{R}^K$, we use $\|f
\|_{T,\infty} = \sup_{a \in \mathcal{A}}\|T(c) f (a) \|_2$. We make
functions in $\Upsilon$ as well as estimators $\widehat \vartheta$ take
values in $[0,1]$. This allows us to simplify the notation in what follows. }

{\small We adopt the standard notation in the empirical process literature
(see, e.g., Van der Vaart, 1998), 
\begin{equation*}
{\mathbb{E}_n}[f] = {\mathbb{E}_n}[f(A)] = n^{-1} \sum_{i=1}^n f(A_i),
\end{equation*}
and 
\begin{equation*}
\mathbb{G}_n[f] = \mathbb{G}_n[f(A)]= n^{-1/2} \sum_{i=1}^n ( f(A_i) - {%
\mathbb{E}}_P[f(A)] ).
\end{equation*}
When the function $\widehat f$ is estimated, the notation should interpreted
as: 
\begin{equation*}
\mathbb{G}_n[\widehat f \ ] = \mathbb{G}_n[f]\mid_{f = \widehat f} \text{\
and \ } {\mathbb{E}}_P[\widehat f \ ] = {\mathbb{E}}_P[f]\mid_{f = \widehat
f}.
\end{equation*}
We also use the concepts of covering entropy and bracketing entropy in the
proofs. The covering entropy $\log N (\epsilon, \mathcal{F}, \|\cdot\|)$ is
the logarithm of the minimal number of $\|\cdot\|$-balls of radius $\epsilon$
needed to cover the set of functions $\mathcal{F}$. The bracketing entropy $%
\log N_{[ ]} (\epsilon, \mathcal{F}, \|\cdot\|)$ is the logarithm of the
minimal number of $\epsilon$-brackets in $\|\cdot\|$ needed to cover the set
of functions $\mathcal{F}$. An $\epsilon$-bracket $[\ell,u]$ in $\|\cdot\|$
is the set of functions $f$ with $\ell \leq f \leq u$ and $\|u - \ell \|<
\epsilon$. }

{\small We follow the notation and definitions in Van der Vaart and Wellner
(1996) of bootstrap consistency. Let $D_{n}$ denote the data vector and $%
E_{n}$ be the vector of bootstrap weights. Consider the random element $%
Z^{b}_{n} = Z_{n}(D_{n}, E_{n})$ in a normed space $\mathbb{Z}$. We say that
the bootstrap law of $Z^{b}_{n}$ consistently estimates the law of some
tight random element $Z$ and write $Z^{b}_{n} \rightsquigarrow_{\Pr} Z $ in $%
\mathbb{Z}$ if 
\begin{equation}  \label{boot1}
\begin{array}{r}
\sup_{h \in\text{BL}_{1}(\mathbb{Z})} \left| {\mathbb{E}}_P^b h \left(
Z^{b}_{n}\right) - {\mathbb{E}}_P h(Z)\right| \rightarrow_{\Pr^b} 0,%
\end{array}%
\end{equation}
where $\text{BL}_{1}(\mathbb{Z})$ denotes the space of functions with the
Lipschitz norm at most 1, ${\mathbb{E}}_P^b$ denotes the conditional
expectation with respect to $E_{n}$ given the data $D_{n}$, and $%
\rightarrow_{\Pr^b}$ denotes convergence in (outer) probability. }

\subsection{{\protect\small Proof of Lemma \protect\ref{thm:fclt}}}

{\small The proof strategy closely follows the argument put forth in
Chernozhukov et al. (2015) to deal with the dimensionality and entropy
properties of the first-step distribution regression estimators. }

\subsubsection{\protect\small Auxiliary Lemmas}

{\small We start with two results on stochastic equicontinuity and a local
expansion for the second-stage estimators that will be used in the proof of
Lemma \ref{thm:fclt}. }

\begin{lemma}[Stochastic equicontinuity]
{\small \label{lemma SE} Let $\omega \geq 0$ be a positive random variable
with ${\mathrm{E}}_P[\omega] = 1$, $\mathrm{Var}_P[\omega] = 1,$ and ${%
\mathbb{E}}_P |\omega|^{2+\delta} < \infty$ for some $\delta > 0$, that is
independent of $(Y*1(C>0),Z,C,V)$, including as a special case $\omega=1$,
and set, for $A = (\omega,Y*1(C>0),Z,C,V)$, 
\begin{equation*}
f_1(A, \vartheta, \beta) := \omega \cdot [W(\vartheta)^{\mathrm{T}}\beta -
Y] \cdot W(\vartheta) \cdot T.
\end{equation*}
Under Assumptions \ref{ass:sampling}--\ref{ass:second}, the following
relations are true. }

\begin{itemize}
\item[(a)] {\small Consider the set of functions 
\begin{equation*}
\mathcal{F} = \{ f_1(A, \vartheta, \beta)^{\mathrm{T}}\alpha :
(\vartheta,\beta) \in \Upsilon_0 \times \mathcal{B}, \alpha \in \mathbb{R}%
^{\dim(W)}, \| \alpha\|_2 \leq 1 \},
\end{equation*}
where $\mathcal{B}$ is a compact set under the $\|\cdot\|_2$ metric
containing $\beta_0$, $\Upsilon_0$ is the intersection of $\Upsilon$,
defined in Lemma \ref{lemma:first}, with a neighborhood of $\vartheta_0$
under the $\|\cdot\|_{T,\infty}$ metric. 
This class is $P$-Donsker with a square integrable envelope of the form $%
\omega$ times a constant. }

\item[(b)] {\small Moreover, if $(\vartheta, \beta) \to (\vartheta_0,
\beta_0) $ in the $\|\cdot \|_{T,\infty} \vee \|\cdot\|_2 $ metric, then 
\begin{equation*}
\| f_1 (A, \vartheta, \beta) - f_1(A, \vartheta_0, \beta_0) \|_{P,2}\to 0.
\end{equation*}
}

\item[(c)] {\small Hence for any $(\widetilde \vartheta, \widetilde \beta)
\to_{\Pr} (\vartheta_0, \beta_0) $ in the $\|\cdot \|_{T,\infty} \vee
\|\cdot\|_2$ metric such that $\widetilde \vartheta \in \Upsilon_0$, 
\begin{equation*}
\|\mathbb{G}_n f_1 ( A,\widetilde \vartheta, \widetilde \beta) - \mathbb{G}%
_n f_1 (A,\vartheta_0, \beta_0) \|_2 \overset{\Pr}{\rightarrow} 0.
\end{equation*}
}

\item[(d)] {\small For for any $(\widehat \vartheta, \widetilde \beta)
\to_{\Pr} (\vartheta_0, \beta_0) $ in the $\|\cdot \|_{T,\infty} \vee
\|\cdot\|_2$ metric so that 
\begin{equation*}
\| \widehat \vartheta - \widetilde \vartheta \|_{T,\infty} = o_\Pr(1/\sqrt{n}%
), \text{ where } \widetilde \vartheta \in \Upsilon_0,
\end{equation*}
we have that 
\begin{equation*}
\|\mathbb{G}_n f_1 (A,\widehat \vartheta, \widetilde \beta) - \mathbb{G}_n
f_1 (A,\vartheta_0, \beta_0)\|_2\to_\Pr 0.
\end{equation*}
}
\end{itemize}
\end{lemma}

{\small \noindent \textbf{Proof of Lemma \ref{lemma SE}.} The proof is
divided into subproofs of each of the claims. 
%
}

{\small \noindent \textbf{Proof of Claim (a)}. The proof proceeds in several
steps. }

{\small \noindent \emph{Step 1}. Here we bound the bracketing entropy for 
\begin{equation*}
\mathcal{I}_1 = \{ [ W(\vartheta)^{\mathrm{T}}\beta - Y] T : \beta \in 
\mathcal{B}, \vartheta \in \Upsilon_0 \}.
\end{equation*}
For this purpose, consider a mesh $\{\vartheta_k\}$ over $\Upsilon_0$ of $%
\|\cdot\|_{T,\infty}$ width $\delta$, and a mesh $\{\beta_l\}$ over $%
\mathcal{B}$ of $\|\cdot\|_2$ width $\delta$. A generic bracket over $%
\mathcal{I}_1$ takes the form 
\begin{equation*}
[i_1^0, i_1^1] = [ \{ W(\vartheta_k)^{\mathrm{T}}\beta_l - \kappa \delta -
Y\} T , \{ W(\vartheta_k)^{\mathrm{T}}\beta_l + \kappa \delta - Y \} T ],
\end{equation*}
where $\kappa = L_W \max_{\beta \in \mathcal{B}} \| \beta\|_2 + L_W $, and $%
L_W := \| \partial_v w \|_{T,\infty} \vee \| w \|_{T, \infty}.$ }

{\small Note that this is a valid bracket for all elements of $\mathcal{I}_1$
because for any $\vartheta$ located within $\delta$ from $\vartheta_k$ and
any $\beta$ located within $\delta$ from $\beta_l$, 
\begin{eqnarray}
|W(\vartheta)^{\mathrm{T}}\beta - W(\vartheta_k)^{\mathrm{T}}\beta_l|T &
\leq & | (W(\vartheta) - W(\vartheta_k))^{\mathrm{T}}\beta | T + |
W(\vartheta_k)^{\mathrm{T}}(\beta - \beta_l)| T  \notag \\
& \leq & L_W \delta \max_{\beta \in \mathcal{B}} \| \beta\|_2 + L_W \delta
\leq \kappa \delta,  \label{eq: converge 1}
\end{eqnarray}
and the $\| \cdot \|_{P,2}$-size of this bracket is given by 
\begin{eqnarray*}
\| i_1^0 - i_1^1 \|_{P,2} & \leq & \sqrt{2} \kappa \delta.
\end{eqnarray*}
Hence, counting the number of brackets induced by the mesh created above, we
arrive at the following relationship between the bracketing entropy of $%
\mathcal{I}_1$ and the covering entropies of $\Upsilon_0$, and $\mathcal{B}$%
, 
\begin{multline*}
\log N_{[]}(\epsilon, \mathcal{I}_1, \|\cdot\|_{P,2}) \lesssim \log
N(\epsilon^2, \Upsilon_0, \|\cdot\|_{T,\infty}) + \log N(\epsilon^2, 
\mathcal{B}, \|\cdot\|_2) \\
\lesssim 1/(\epsilon^2 \log^4 \epsilon) + \log(1/\epsilon),
\end{multline*}
and so $\mathcal{I}_1$ is $P$-Donsker with a constant envelope. }

{\small \noindent \emph{Step 2}. Similarly to Step 1, it follows that 
\begin{equation*}
\mathcal{I}_2 = \{ W(\vartheta)^{\mathrm{T}}\alpha T : \vartheta \in
\Upsilon_0, \alpha \in \mathbb{R}^{\dim(W)}, \|\alpha\|_2 \leq 1 \}
\end{equation*}
also obeys a similar bracketing entropy bound 
\begin{equation*}
\log N_{[]}(\epsilon, \|\cdot\|_{P,2}) \lesssim 1/(\epsilon^2 \log^4
\epsilon) + \log(1/\epsilon)
\end{equation*}
with a generic bracket taking the form $[i_2^0,i_2^1] = [\{W(\vartheta_k)^{%
\mathrm{T}}\alpha - \kappa \delta\}T, \{W(\vartheta_k)^{\mathrm{T}}\alpha +
\kappa \delta\}T]$. Hence, this class is also $P$-Donsker with a constant
envelope. }

{\small \noindent \emph{Step 3}. In this step, we verify the claim (a). Note
that $\mathcal{F} = \omega \cdot \mathcal{I}_1 \cdot \mathcal{I}_2. $ This
class has a square-integrable envelope under $P$. 
The class $\mathcal{F}$ is $P$-Donsker by the following argument. Note that
the product $\mathcal{I}_1 \cdot \mathcal{I}_2$ of uniformly bounded classes
is $P$-Donsker, e.g., by Theorem 2.10.6 of Van der Vaart and Wellner (1996).
Under the stated assumption the final product of the random variable $\omega$
with the $P$-Donsker class remains to be $P$-Donsker by the Multiplier
Donsker Theorem, namely Theorem 2.9.2 in Van der Vaart and Wellner (1996). 
\newline
}

{\small \noindent \textbf{Proof of Claim (b)}. The claim follows by the
Dominated Convergence Theorem, since any $f_1 \in \mathcal{F}$ is dominated
by a square-integrable envelope under $P$ and $W(\vartheta)^{\mathrm{T}%
}\beta T \to W^{\mathrm{T}}\beta_0 T $ and $|W(\vartheta)^{\mathrm{T}}\alpha
T - W^{\mathrm{T}}\alpha T| \to 0$ in view of the relation such as (\ref{eq:
converge 1}). \newline
}

{\small \noindent \textbf{Proof of Claim (c)}. This claim follows from the
asymptotic equicontinuity of the empirical process $(\mathbb{G}_n [f_1], f_1
\in \mathcal{F})$ under the $L_2(P)$ metric, and hence also with respect to
the $\|\cdot\|_{T,\infty} \vee \|\cdot\|_2$ metric in view of Claim (b). 
\newline
}

{\small \noindent \textbf{Proof of Claim (d)}. It is convenient to set $%
\widehat f_1 := f_1 (A, \widehat \vartheta, \widetilde \beta)$ and $%
\widetilde f_1 := f_1 (A, \widetilde \vartheta, \widetilde \beta).$ Note
that 
\begin{eqnarray*}
\max_{1\leq j \leq \dim W} | \mathbb{G}_n [\widehat f_1 - \widetilde f_1]|_j
& \leq & \max_{1\leq j \leq \dim W} |\sqrt{n} {\mathbb{E}_n} [\widehat f_1 -
\widetilde f_1 ]|_j + \max_{1\leq j \leq \dim W} |\sqrt{n} {\mathbb{E}}_P
(\widehat f_1 - \widetilde f_1)|_j \\
& \lesssim & \sqrt{n} {\mathbb{E}_n}[ \widehat \zeta \ ] + \sqrt{n} {\mathrm{%
E}}_P [\widehat \zeta \ ] \\
&\lesssim& \mathbb{G}_n [\widehat \zeta \ ] + 2 \sqrt{n} {\mathbb{E}}_P
[\widehat \zeta \ ],
\end{eqnarray*}
where $|f|_j$ denotes the $j$th element of the application of absolute value
to each element of the vector $f_1$, and $\widehat \zeta$ is defined by the
following relationship, which holds with probability approaching one, 
\begin{equation*}  \label{eq: bound diff}
\max_{1\leq j \leq \dim W} |\widehat f_1 - \widetilde f_1 |_j \lesssim
\omega |W(\widehat \vartheta)^{\mathrm{T}}\widetilde \beta - Y| \|W(\widehat
\vartheta) - W(\widetilde \vartheta)\|_2 + \omega |W(\widehat \vartheta)^{%
\mathrm{T}}\widetilde \beta - W(\widetilde \vartheta)^{\mathrm{T}}\widetilde
\beta| \lesssim \omega (k + |Y|) \Delta_n =:\widehat \zeta,
\end{equation*}
where $k$ is a constant such that $k \geq L_W \max_{\beta \in \mathcal{B}}
\|\beta\|_2$ with $L_W = \| \partial_v w \|_{T,\infty} \vee \| w
\|_{T,\infty}$, and $\Delta_n = o (1/\sqrt{n})$ is a deterministic sequence
such that 
\begin{equation*}
\Delta_n \geq\| \widehat \vartheta - \widetilde \vartheta\|_{T,\infty}.
\end{equation*}
The second inequality result follows from 
\begin{equation*}
|W(\widehat \vartheta)^{\mathrm{T}}\widetilde \beta - Y| \|W(\widehat
\vartheta) - W(\widetilde \vartheta)\|_2 \lesssim (k + |Y|) \Delta_n, \text{
and } |W(\widehat \vartheta)^{\mathrm{T}}\widetilde \beta - W(\widetilde
\vartheta)^{\mathrm{T}}\widetilde \beta| \lesssim k \Delta_n.
\end{equation*}
Then, by part (c) the result follows from 
\begin{eqnarray*}
\mathbb{G}_n [ \widehat \zeta \ ]= o_\Pr(1), \ \ \ \ \sqrt{n} {\mathbb{E}}_P
[ \widehat \zeta \ ] = o_\Pr(1).
\end{eqnarray*}
Indeed, 
\begin{equation*}
\| \widehat \zeta \|_{P,2} \lesssim \sqrt{{\mathbb{E}}_P \omega^2 (k^2 + {%
\mathrm{E}}_P (Y^2 \mid C \in \overline{\mathcal{C}})) \Delta_n^2}= o(1)
\Rightarrow \mathbb{G}_n [\widehat \zeta \ ] = o_\Pr(1),
\end{equation*}
and 
\begin{equation*}
\| \widehat \zeta \|_{P,1} \leq {\mathbb{E}}_P|\omega| \cdot (k + {\mathbb{E}%
}_P( |Y| \mid C \in \overline{\mathcal{C}}) \Delta_n = o(1/\sqrt{n})
\Rightarrow {\mathbb{E}}_P |\widehat \zeta| = o_\Pr(1/\sqrt{n}).
\end{equation*}
}

\begin{lemma}[Local expansion]
{\small \label{lemma Expand} Under Assumptions \ref{ass:sampling}--\ref%
{ass:second}, for 
\begin{eqnarray*}
&& \widehat \delta = \sqrt{n}(\widetilde \beta - \beta_0) = O_{\Pr}(1); \\
&& \widehat \Delta(c,r) = \sqrt{n}(\widehat \vartheta(c,r)-
\vartheta_0(c,r)) = \sqrt{n} \ {\mathbb{E}_n}[\ell(A, c,r)] + o_{\Pr}(1) 
\text{ in } \ell^{\infty}(\overline{\mathcal{CR}}), \\
&& \|\sqrt{n} \ {\mathbb{E}_n}[\ell(A,\cdot)]\|_{T, \infty} =O_{\Pr}(1),
\end{eqnarray*}
we have that 
\begin{eqnarray}
& & \sqrt{n} \ {\mathbb{E}}_P [\{W(\widehat \vartheta)^{\mathrm{T}%
}\widetilde \beta - Y \} W(\widehat \vartheta) T] = J \widehat \delta + 
\sqrt{n} \ {\mathbb{E}_n} \left[ f_2(A) \right] + o_{\Pr}(1),  \notag
\end{eqnarray}
where 
\begin{equation*}
f_2(a) = {\mathbb{E}}_P \{ [ W^{\mathrm{T}}\beta_0 - Y] \dot W + W \dot W^{%
\mathrm{T}}\beta_0 \} T \ell(a, C,R).
\end{equation*}
}
\end{lemma}

{\small \noindent \textbf{Proof of Lemma \ref{lemma Expand}.} }

{\small \noindent Uniformly in $Z \in \overline{\mathcal{Z}}$, 
\begin{eqnarray*}
& & \sqrt{n} {\mathbb{E}}_P\{ W(\widehat \vartheta)^{\mathrm{T}}\widetilde
\beta - Y \mid Z\} T  \notag \\
& &= \sqrt{n} {\mathbb{E}}_P\{ W^{\mathrm{T}}\beta_0 - Y \mid Z\} T + \{ W(%
\bar{\vartheta}_{\xi})^{\mathrm{T}}\widehat \delta + \dot{W}(\bar{\vartheta}%
_{\xi})^{\mathrm{T}} \bar \beta_{\xi} \widehat \Delta(C,R) \} T \\
& & = \sqrt{n} {\mathbb{E}}_P\{ W^{\mathrm{T}}\beta_0 - Y \mid Z\} T + \{ W^{%
\mathrm{T}}\widehat \delta + \dot W^{\mathrm{T}}\beta_0 \widehat \Delta
(C,R)\} T + \rho_{Z}, \\
& & \bar \rho = \sup_{\{Z \in \overline{\mathcal{Z}}\}} |\rho_{Z}| =
o_{\Pr}(1),  \label{eq: expand 2}
\end{eqnarray*}
where $\bar{\vartheta}_{\xi}$ is on the line connecting $\vartheta_0$ and $%
\widehat \vartheta$ and $\bar \beta_{\xi}$ is on the line connecting $%
\beta_0 $ and $\widetilde \beta$. The first equality follows by the mean
value expansion. The second equality follows by uniform continuity of $%
W(\cdot)$ and $\dot{W}(\cdot)$, $\|\widehat \vartheta - \vartheta_0\|_{T,
\infty} \overset{\Pr}{\rightarrow} 0$ and $\|\widetilde \beta - \beta_0\|_2 
\overset{\Pr}{\rightarrow} 0$. }

{\small Since the entries of $W$ and $\dot W$ are bounded, $\widehat \delta
= O_\Pr(1),$ and $\|\widehat \Delta\|_{T, \infty} = O_\Pr(1)$, with
probability approaching one, 
\begin{multline*}  \label{eq: exp1}
\sqrt{n} {\mathbb{E}}_P \{W(\widehat \vartheta)^{\mathrm{T}}\widetilde \beta
- Y \} W(\widehat \vartheta) T = {\mathbb{E}}_P \{W^{\mathrm{T}}\beta_0 - Y
\} \dot W T \widehat \Delta (C,R) \\
+ {\mathbb{E}}_P\{W W^{\mathrm{T}} T \} \widehat \delta + {\mathbb{E}}_P\{W
\dot W^{\mathrm{T}}\beta_0 T \widehat \Delta (C,R) \}+ O_{\Pr}(\bar \rho) \\
= J \widehat \delta + {\mathbb{E}}_P[ \{W^{\mathrm{T}}\beta_0 - Y \} \dot W
+ W \dot W^{\mathrm{T}}\beta_0] T \widehat \Delta(C,R) + o_{\Pr} (1).
\end{multline*}
Substituting in $\widehat \Delta(x,r) = \sqrt{n} \ {\mathbb{E}_n}[ \ell(A,
x,r)] + o_{\Pr}(1)$ and interchanging ${\mathbb{E}}_P$ and ${\mathbb{E}_n}$, 
\begin{equation*}
{\mathbb{E}}_P [ \{W^{\mathrm{T}}\beta_0 - Y \} \dot W + W \dot W^{\mathrm{T}%
}\beta_0] T \widehat \Delta(C,R) = \sqrt{n} \ {\mathbb{E}_n}[ g(A)] +
o_{\Pr}(1),
\end{equation*}
since $\|[ \{W^{\mathrm{T}}\beta_0 - Y \} \dot W + \ W \dot W^{\mathrm{T}%
}\beta_0]T\|_{P,2}$ is bounded . The claim of the lemma follows. }

\subsubsection{{\protect\small Proof of Lemma \protect\ref{thm:fclt}.}}

{\small The proof is divided in two parts corresponding to the CLT and
bootstrap CLT. }

\paragraph{\protect\small CLT:}

{\small In this part we show $\sqrt{n}(\widehat\beta - \beta_0)
\rightsquigarrow J^{-1} G$ in $\mathbb{R}^{d_w}$. }

{\small \noindent \emph{Step 1}. This step shows that $\sqrt{n}(\widehat
\beta - \beta_0) = O_\Pr(1)$. Recall that 
\begin{equation*}
\widehat \beta = \arg \min_{\beta \in \mathbb{R}^{d_w}} \mathbb{E}_n [(Y -
W(\widehat \vartheta)^{\mathrm{T}}\beta)^2 T].
\end{equation*}
Due to the convexity of the objective function, it suffices to show that for
any $\epsilon>0$, there exists a finite positive constant $B_{\epsilon}$
such that, 
\begin{eqnarray}  \label{eq: prob0}
\liminf_{n \to \infty } \Pr \left( \inf_{\|\eta\|_2=1}\sqrt{n} \eta^{\mathrm{%
T}} {\mathbb{E}_n} \Big [\widehat f_{1, \eta,B_{\epsilon}} \Big ] > 0
\right) \geq 1- \epsilon,
\end{eqnarray}
where 
\begin{equation*}
\widehat f_{1, \eta,B_{\epsilon}} (A) :=\left\{W(\widehat \vartheta)^{%
\mathrm{T}}(\beta_0 + B_\epsilon \eta/\sqrt{n}) - Y \right\}
W(\widehat\vartheta) T.
\end{equation*}
Let 
\begin{equation*}
f_1(A): = \left\{W^{\mathrm{T}}\beta_0 - Y \right\} W T.
\end{equation*}
Then uniformly in $\|\eta\|_2=1$, 
\begin{eqnarray*}
\sqrt{n} \eta^{\mathrm{T}} {\mathbb{E}_n}[ \widehat f_{1,
\eta,B_{\epsilon}}] & = & \eta^{\mathrm{T}} \mathbb{G}_n[ \widehat f_{1,
\eta,B_{\epsilon}}] + \sqrt{n} \eta^{\mathrm{T}} {\mathbb{E}}_P[ \widehat
f_{1, \eta,B_{\epsilon}}] \\
& =_{(1)} & \eta^{\mathrm{T}}\mathbb{G}_n [f_1] + o_\Pr(1) + \eta^{\mathrm{T}%
} \sqrt{n} {\mathbb{E}}_P[ \widehat f_{1, \eta,B_{\epsilon}}] \\
& =_{(2)} & \eta^{\mathrm{T}} \mathbb{G}_n [f_1] + o_\Pr(1) + \eta^{\mathrm{T%
}} J\eta B_{\epsilon} + \eta^{\mathrm{T}}\mathbb{G}_n[f_2] + o_\Pr(1) \\
& =_{(3)} & O_\Pr(1) + o_\Pr(1) + \eta^{\mathrm{T}} J\eta B_{\epsilon} +
O_\Pr(1) + o_\Pr(1),
\end{eqnarray*}
where the relations (1) and (2) follow by Lemma \ref{lemma SE} and Lemma \ref%
{lemma Expand} with $\widetilde \beta = \beta_0 + B_\epsilon \eta/\sqrt{n}$,
respectively, using the fact that $\|\widehat \vartheta - \widetilde
\vartheta \|_{T,\infty} = o_\Pr(1/\sqrt{n})$, $\widetilde \vartheta \in
\Upsilon$, $\|\widetilde \vartheta - \vartheta_0\|_{T,\infty} = O_\Pr (1/%
\sqrt{n})$ and $\| \beta_0 + B_\epsilon \eta/\sqrt{n} - \beta_0 \|_2 = O(1/%
\sqrt{n})$; relation (3) holds because $f_1$ and $f_2$ are $P$-Donsker by
step-2 below. Since $J$ is positive definite, with minimal eigenvalue
bounded away from zero, the inequality (\ref{eq: prob0}) follows by choosing 
$B_{\epsilon}$ as a sufficiently large constant. }

{\small \noindent \emph{Step 2}. In this step we show the main result. Let 
\begin{equation*}
\widehat f_1(A) := \left\{ W(\widehat \vartheta)^{\mathrm{T}}\widehat \beta
- Y \right\} W(\widehat\vartheta) T.
\end{equation*}
From the first-order conditions of the least squares problem, 
\begin{eqnarray*}
0 = \sqrt{n} {\mathbb{E}_n}\left[ \widehat f_1 \right] & = & \mathbb{G}_n%
\left[ \widehat f_1 \right] + \sqrt{n} {\mathbb{E}}_P\left[ \widehat f_1 %
\right] \\
& =_{(1)} & \mathbb{G}_n [f_1] + o_\Pr(1) + \sqrt{n} {\mathbb{E}}_P\left[
\widehat f_1 \right] \\
& =_{(2)} & \mathbb{G}_n [f_1] + o_\Pr(1) + J \sqrt{n}(\widehat \beta -
\beta_0) + \mathbb{G}_n[f_2]+ o_\Pr(1),
\end{eqnarray*}
where relations (1) and (2) follow by Lemma \ref{lemma SE} and Lemma \ref%
{lemma Expand} with $\widetilde \beta = \widehat \beta$, respectively, using
that $\|\widehat \vartheta - \widetilde \vartheta\|_{T,\infty} = o_\Pr(1/%
\sqrt{n})$, $\widetilde \vartheta \in \Upsilon$, and $\|\widetilde \vartheta
- \vartheta\|_{T,\infty} = O_\Pr (1/\sqrt{n})$ by Lemma \ref{lemma:first},
and $\| \widehat \beta - \beta_0\|_2 = O_\Pr(1/\sqrt{n})$. }

{\small Therefore, by the invertibility of $J$, 
\begin{equation*}
\sqrt{n}(\widehat \beta - \beta_0) = - J^{-1} \mathbb{G}_n(f_1 + f_2) +
o_\Pr(1).
\end{equation*}
By the Central Limit Theorem 
\begin{equation*}
\mathbb{G}_n(f_1 + f_2) \rightsquigarrow G \text{ in $\mathbb{R}^{d_w}$}, \
\ G \sim N(0,\Omega), \ \ \Omega = {\mathbb{E}}_P[(f_1 + f_2)(f_1 + f_2)^{%
\mathrm{T}}],
\end{equation*}
where $\Omega$ is specified in the lemma. Conclude that 
\begin{equation*}
\sqrt{n}(\widehat \beta - \beta_0) \rightsquigarrow J^{-1} G \text{ in $%
\mathbb{R}^{d_w}$}.
\end{equation*}
}

{\small 
}

\paragraph{\protect\small Bootstrap CLT:}

{\small In this part we show $\sqrt{n}(\widehat\beta^b - \widehat \beta)
\rightsquigarrow_{\Pr} J^{-1} G$ in $\mathbb{R}^{d_w}$. }

{\small \noindent \emph{Step 1}. This step shows that $\sqrt{n}(\widehat
\beta^b - \beta_0) = O_\Pr(1) $ under the unconditional probability $\Pr$.
Recall that 
\begin{equation*}
\widehat \beta^b = \arg \min_{\beta \in \mathbb{R}^{\dim(W)}} \mathbb{E}_n
[\omega (Y - W(\widehat \vartheta^b)^{\mathrm{T}}\beta)^2 T ],
\end{equation*}
where $\omega$ is the random variable used in the weighted bootstrap. 
Due to convexity of the objective function, it suffices to show that for any 
$\epsilon>0$ there exists a finite positive constant $B_{\epsilon}$ such
that 
\begin{eqnarray}  \label{eq: prob}
\liminf_{n \to \infty } \Pr \left( \inf_{\|\eta\|_2=1}\sqrt{n} \eta^{\mathrm{%
T}} {\mathbb{E}_n} \Big [\widehat f^b_{1, \eta,B_{\epsilon}} \Big ] > 0
\right) \geq 1- \epsilon,
\end{eqnarray}
where 
\begin{equation*}
\widehat f^b_{1, \eta,B_{\epsilon}} (A) := \omega \cdot \left\{ [W(\widehat
\vartheta^b)^{\mathrm{T}}(\beta_0 + B_\epsilon \eta/\sqrt{n})] - Y\right\}
W(\widehat\vartheta^b) T.
\end{equation*}
The result then follows by an analogous argument to step 1 in the proof of
the CLT, which we do not repeat here. }

{\small 
}

{\small \noindent \emph{Step 2}. In this step, we show that $\sqrt{n}%
(\widehat \beta^b - \beta_0) = - J^{-1} \mathbb{G}_n( f_1^b + f_2^b) +
o_\Pr(1)$ under the unconditional probability $\Pr$. Let 
\begin{equation*}
\widehat f_1^b(A) := \omega \cdot \{ W(\widehat \vartheta^b)^{\mathrm{T}%
}\widehat \beta^b - Y \} W(\widehat\vartheta^b) T.
\end{equation*}
From the first-order conditions of the least squares problem in the weighted
sample, 
\begin{eqnarray*}
0 = \sqrt{n} {\mathbb{E}_n}\left[ \widehat f_1^b \right] & = & \mathbb{G}_n%
\left[ \widehat f_1^b \right] + \sqrt{n} {\mathbb{E}}_{P}\left[ \widehat
f_1^b \right] \\
& =_{(1)} & \mathbb{G}_n [f_1^b] + o_\Pr(1) + \sqrt{n} {\mathbb{E}}_{P}\left[
\widehat f_1^b \right] \\
& =_{(2)} & \mathbb{G}_n [f_1^b] + o_\Pr(1) + J \sqrt{n}(\widehat \beta^b -
\beta_0) + \mathbb{G}_n[f_2^b]+ o_\Pr(1),
\end{eqnarray*}
where the relations (1) and (2) follow by Lemma \ref{lemma SE} and Lemma \ref%
{lemma Expand} with $\widetilde \beta = \widehat \beta^b $, respectively,
using the fact that $\|\widehat \vartheta^b - \widetilde
\vartheta^b\|_{T,\infty} = o_\Pr(1/\sqrt{n})$, $\widetilde \vartheta^b \in
\Upsilon$ and $\|\widetilde \vartheta^b - \vartheta_0\|_{T,\infty} = O_\Pr
(1/\sqrt{n})$ by Lemma \ref{lemma:first}, and $\| \widehat \beta^b -
\beta_0\|_2 = O_\Pr(1/\sqrt{n})$. Therefore, by the invertibility of $J$, 
\begin{equation*}
\sqrt{n}(\widehat \beta^b - \beta_0) = - J^{-1} \mathbb{G}_n(f_1^b + f_2^b)
+ o_\Pr(1).
\end{equation*}
}

{\small \noindent \emph{Step 3}. In this final step we establish the
behavior of $\sqrt{n}(\widehat \beta^b - \widehat \beta)$ under $\Pr^b$.
Note that $\Pr^b$ denotes the conditional probability measure, namely the
probability measure induced by draws of $\omega_1,\ldots,\omega_n$
conditional on the data $A_1,...,A_n$. By Step 2 of the proof of the CLT and
Step 2 of the proof of the bootstrap CLT, we have that under $\Pr$: 
\begin{equation*}
\sqrt{n}(\widehat \beta^b - \beta_0) = - J^{-1} \mathbb{G}_n(f_1^b + f_2^b)
+ o_\Pr(1), \ \sqrt{n}(\widehat \beta - \beta_0) = - J^{-1} \mathbb{G}_n(f_1
+ f_2) + o_\Pr(1).
\end{equation*}
Hence, under $\Pr$ 
\begin{equation*}
\sqrt{n}(\widehat \beta^b - \widehat \beta) = - J^{-1} \mathbb{G}_n(f_1^b
-f_1 + f_2^b - f_2) + r_n = - J^{-1} \mathbb{G}_n( (\omega-1) (f_1 + f_2)) +
r_n ,
\end{equation*}
where $r_n = o_\Pr(1)$. Note that it is also true that 
\begin{equation*}
r_n = o_{\Pr^b}(1) \text{ in $\Pr$-probability},
\end{equation*}
where the latter statement means that for every $\epsilon>0$, $\Pr^b( \|r_n
\|_2 > \epsilon) = {o}_{\Pr}(1). $ Indeed, this follows from Markov
inequality and by 
\begin{equation*}
{\mathbb{E}}_{\mathbb{P}}[ \Pr^b( \|r_n \|_2 > \epsilon) ] = \Pr(\|r_n \|_2
> \epsilon ) = o(1),
\end{equation*}
where the latter holds by the Law of Iterated Expectations and $r_n =
o_\Pr(1)$. }

{\small Note that $f_1^b = \omega \cdot f_1$ and $f_2^b = \omega \cdot f_2$,
where $f_1$ and $f_2$ are $P$-Donsker by step-2 of the proof of the first
part and ${\mathbb{E}}_P \omega^2 < \infty$. Then, by the Conditional
Multiplier Central Limit Theorem, e.g., Lemma 2.9.5 in Van der Vaart and
Wellner (1996), 
\begin{equation*}
G^b_n := \mathbb{G}_n( (\omega-1) (f_1 + f_2)) \rightsquigarrow_{\Pr} G 
\text{ in }\mathbb{R}^{d_w}.
\end{equation*}
Conclude that 
\begin{equation*}
\sqrt{n}(\widehat \beta^b - \widehat \beta) \rightsquigarrow_{\Pr} J^{-1} G 
\text{ in } \mathbb{R}^{d_w}.
\end{equation*}
}

\subsection{{\protect\small Proof of Theorem \protect\ref{fclt:sdf}}}

{\small In this section we use the notation $W_x(\vartheta) =
w(x,Z_1,V(\vartheta))$ such that $W_x = w(x,Z_1,V(\vartheta_0))$. }

{\small We focus on the proof for the estimator of the ASF, because the
proof for the estimator of the ASF on the treated can be obtained by
analogous arguments. The results for the estimator of the ASF follow by a
similar argument to the proof of Lemma \ref{thm:fclt} using Lemmas \ref%
{lemma SE2} and \ref{lemma Expand2} in place of Lemmas \ref{lemma SE} and %
\ref{lemma Expand}, and the delta method. For the sake of brevity, here we
just outline the proof of the FCLT. }

{\small Let $\psi_{x}(A,\vartheta,\beta):= W_{x}(\vartheta)^{\mathrm{T}}
\beta T$ such that $\mu_S(x)=$ ${\mathbb{E}}_{P}\psi_{x}(A,\vartheta_{0},%
\beta_{0})/{\mathbb{E}}_P T$ and $\widehat{\mu}_S(x)={\mathbb{E}_n}%
\psi_{x}(A,\widehat{\vartheta},\widehat{\beta})/{\mathbb{E}_n} T$. Then, for 
$\widehat{\psi}_{x}:=\psi_{x}(A,\widehat{\vartheta},\widehat{\beta})$ and $%
\psi_{x}:=\psi_{x}(A,\vartheta_{0},\beta_{0})$, 
\begin{eqnarray*}
\sqrt{n}\left[{\mathbb{E}_n}\psi_{x}(A,\widehat{\vartheta},\widehat{\beta})-{%
\mathbb{E}}_{P}\psi_{x}(A,\vartheta_{0},\beta_{0})\right] & = & \mathbb{G}_n%
\left[\widehat{\psi}_{x}\right]+\sqrt{n}{\mathbb{E}}_{P}\left[\widehat{\psi}%
_{x}-\psi_{x}\right] \\
& =_{(1)} & \mathbb{G}_n[\psi_{x}]+{o}_{\Pr}(1)+\sqrt{n}{\mathbb{E}}_{P}%
\left[\widehat{\psi}_{x}-\psi_{x}\right] \\
& =_{(2)} & \mathbb{G}_n[\psi_{x}]+{o}_{\Pr}(1)+\mathbb{G}_n[\sigma_{x}]+{o}%
_{\Pr}(1),
\end{eqnarray*}
where relations (1) and (2) follow by Lemma \ref{lemma SE2} and Lemma \ref%
{lemma Expand2} with $\widetilde{\beta}=\widehat{\beta}$, respectively,
using that $\|\widehat{\vartheta}-\widetilde{\vartheta}\|_{T,\infty}=o_{%
\Pr}(1/\sqrt{n})$, $\widetilde{\vartheta}\in \Upsilon$, and $\|\widetilde{%
\vartheta}-\vartheta\|_{T,\infty}=O_{\Pr}(1/\sqrt{n})$ by Lemma \ref%
{lemma:first}, and $\sqrt{n}(\widehat{\beta}-\beta_{0})=-J^{-1}\mathbb{G}%
_n(f_1+f_2)+o_{\Pr}(1)$ from step 2 of the proof of Lemma \ref{thm:fclt}. }

{\small The functions $x \mapsto\psi_{x}$ and $x \mapsto \sigma_{x}$ are $P$%
-Donsker by Example 19.7 in Van der Vaart (1998) because they are Lipschitz
continuous on $\overline{\mathcal{X}}$. Hence, by the Functional Central
Limit Theorem 
\begin{equation*}
\mathbb{G}_n(\psi_{x}+\sigma_{x})\rightsquigarrow Z(x)\text{ in }%
\ell^{\infty}(\overline{\mathcal{X}}),
\end{equation*}
where $x \mapsto Z(x)$ is a zero mean Gaussian process with uniformly
continuous sample paths and covariance function 
\begin{equation*}
\mathrm{Cov}_{P}[\psi_{x_0}+\sigma_{x_0},\psi_{x_1}+\sigma_{x_1}],\ \
x_0,x_1 \in\overline{\mathcal{X}}.
\end{equation*}
The result follows by the functional delta method applied to the ratio of ${%
\mathbb{E}_n}\psi_{x}(A,\widehat{\vartheta},\widehat{\beta})$ and ${\mathbb{E%
}_n} T$ using that 
\begin{equation*}
\left( 
\begin{array}{c}
{\mathbb{G}_n}\psi_{x}(A,\widehat{\vartheta},\widehat{\beta}) \\ 
{\mathbb{G}_n} T%
\end{array}
\right) \rightsquigarrow \left( 
\begin{array}{c}
Z(x) \\ 
Z_T%
\end{array}
\right),
\end{equation*}
where $Z_T \sim N(0, p_T(1-p_T))$, 
\begin{equation*}
\mathrm{Cov}_P(Z(x),Z_T) = G_T(x) p_T(1-p_T),
\end{equation*}
and 
\begin{multline*}
\mathrm{Cov}_{P}[\psi_{x_0}+h_{x_0},\psi_{x_1}+\sigma_{x_1} \mid T = 1] \\
= \frac{\mathrm{Cov}_{P}[\psi_{x_0}+\sigma_{x_0},\psi_{x_1}+\sigma_{x_1}] -
\mu_T(x_0) \mu_T(x_1) p_T(1-p_T)}{p_T}.
\end{multline*}
}

\begin{lemma}[Stochastic equicontinuity]
{\small \label{lemma SE2} Let $\omega \geq 0$ be a positive random variable
with ${\mathbb{E}}_P[\omega] = 1$, $\mathrm{Var}_P[\omega] = 1,$ and ${%
\mathbb{E}}_P |\omega|^{2+\delta} < \infty$ for some $\delta > 0$, that is
independent of $(Y 1(C>0),C,Z,V)$, including as a special case $\omega=1$,
and set, for $A = (\omega,Y 1(C>0),C,Z,V)$, 
\begin{equation*}
\psi_x(A, \vartheta, \beta) := \omega \cdot W_x(\vartheta)^{\mathrm{T}}\beta
\cdot T.
\end{equation*}
Under Assumptions \ref{ass:sampling}--\ref{ass:second}, the following
relations are true. }

\begin{itemize}
\item[(a)] {\small Consider the set of functions 
\begin{equation*}
\mathcal{F} := \{ \psi_x(A, \vartheta, \beta) : (\vartheta,\beta,x) \in
\Upsilon_0 \times \mathcal{B} \times \overline{\mathcal{X}} \},
\end{equation*}
where $\overline{\mathcal{X}} $ is a compact subset of $\mathbb{R}$, $%
\mathcal{B}$ is a compact set under the $\|\cdot\|_2$ metric containing $%
\beta_0$, $\Upsilon_0$ is the intersection of $\Upsilon$, defined in Lemma %
\ref{lemma:first}, with a neighborhood of $\vartheta_0$ under the $%
\|\cdot\|_{T,\infty}$ metric. This class is $P$-Donsker with a square
integrable envelope of the form $\omega$ times a constant. }

\item[(b)] {\small Moreover, if $(\vartheta, \beta) \to (\vartheta_0,
\beta_0) $ in the $\|\cdot \|_{T,\infty} \vee \|\cdot\|_2 $ metric , then 
\begin{equation*}
\sup_{x\in\overline{\mathcal{X}}} \| \psi_x (A, \vartheta, \beta) -
\psi_x(A, \vartheta_0, \beta_0) \|_{P,2}\to 0.
\end{equation*}
}

\item[(c)] {\small Hence for any $(\widetilde \vartheta, \widetilde \beta)
\to_{\Pr} (\vartheta_0, \beta_0) $ in the $\|\cdot \|_{T,\infty} \vee
\|\cdot\|_2$ metric such that $\widetilde \vartheta \in \Upsilon_0$, 
\begin{equation*}
\sup_{x\in\overline{\mathcal{X}}} \|\mathbb{G}_n \psi_x ( A,\widetilde
\vartheta, \widetilde \beta) - \mathbb{G}_n \psi_x (A,\vartheta_0, \beta_0)
\|_2 \to_\Pr 0.
\end{equation*}
}

\item[(d)] {\small For for any $(\widehat \vartheta, \widetilde \beta)
\to_{\Pr} (\vartheta_0, \beta_0) $ in the $\|\cdot \|_{T,\infty} \vee
\|\cdot\|_2$ metric , so that 
\begin{equation*}
\| \widehat \vartheta - \widetilde \vartheta \|_{T,\infty} = o_\Pr(1/\sqrt{n}%
), \text{ where } \widetilde \vartheta \in \Upsilon_0,
\end{equation*}
we have that 
\begin{equation*}
\sup_{x\in\overline{\mathcal{X}}} \|\mathbb{G}_n \psi_x (A,\widehat
\vartheta, \widetilde \beta) - \mathbb{G}_n \psi_x (A,\vartheta_0,
\beta_0)\|_2\to_\Pr 0.
\end{equation*}
}
\end{itemize}
\end{lemma}

{\small \noindent \textbf{Proof of Lemma \ref{lemma SE2}.} The proof is
omitted because it is similar to the proof of Lemma \ref{lemma SE}. }

\begin{lemma}[Local expansion]
{\small \label{lemma Expand2} Under Assumptions \ref{ass:sampling}--\ref%
{ass:second}, for 
\begin{eqnarray*}
&& \widehat \delta = \sqrt{n}(\widetilde \beta - \beta_0) = O_{\Pr}(1); \\
&& \widehat \Delta(c,r) = \sqrt{n}(\widehat \vartheta(c,r)-
\vartheta_0(c,r)) = \sqrt{n} \ {\mathbb{E}_n}[\ell(A, c,r)] + o_{\Pr}(1) 
\text{ in } \ell^{\infty}(\overline{\mathcal{CR}}), \\
&& \|\sqrt{n} \ {\mathbb{E}_n}[\ell(A,\cdot)]\|_{T, \infty} =O_{\Pr}(1),
\end{eqnarray*}
we have that 
\begin{multline*}
\sqrt{n} \ \left\{ {\mathbb{E}}_P W_x(\widehat \vartheta)^{\mathrm{T}%
}\widetilde \beta T - {\mathbb{E}}_P W_x^{\mathrm{T}}\beta_0 T \right\} = {%
\mathbb{E}}_P\{ W_x T \}^{\mathrm{T}} \widehat{\delta} + {\mathbb{E}}_P\{
\dot W_x^{\mathrm{T}}\beta_0 T \ell(a, C,R)\} \big |_{a=A} + \bar o_{\Pr}(1),
\end{multline*}
where $\bar o_{\Pr}(1)$ denotes order in probability uniform in $x \in 
\overline{\mathcal{X}}$. }
\end{lemma}

{\small \noindent \textbf{Proof of Lemma \ref{lemma Expand2}.} The proof is
omitted because it is similar to the proof of Lemma \ref{lemma Expand}. }

\end{document}